\documentclass[11pt,a4paper]{article}
\usepackage[margin=1.0in]{geometry}
\usepackage{aas_macros,amsmath,amssymb,esint,hyperref,microtype,subcaption,tikz}

\hypersetup{colorlinks}
\numberwithin{equation}{section}
\usetikzlibrary{arrows,arrows.meta,backgrounds,calc,decorations.markings,decorations.pathmorphing,decorations.pathreplacing,fpu}

\graphicspath{{figs/}}

\let\originalleft\left
\let\originalright\right
\renewcommand{\left}{\mathopen{}\mathclose\bgroup\originalleft}
\renewcommand{\right}{\aftergroup\egroup\originalright}

\newcommand{\br}[1]{\left[#1\right]}
\newcommand{\pa}[1]{\left(#1\right)}
\newcommand{\ed}{\mathop{}\!\mathrm{d}}
\newcommand{\pd}{\mathop{}\!\partial}
\DeclareMathOperator{\Tr}{Tr}

\title{Chaos and fractals of the black hole photon ring}

\author{\normalsize Roman Berens$^{1}$, Peter Galison$^{2}$, Trevor Gravely$^{1}$, Alexandru Lupsasca$^{1,3}$, and Leo C.\ Stein$^{4}$}
\date{\textit{\normalsize $^1$Vanderbilt University\quad $^2$Harvard University\quad $^3$OpenAI\quad $^4$University of Mississippi}}

\begin{document}

\maketitle

\begin{abstract}
    The photon ring of a Kerr black hole decomposes into a self-similar hierarchy of subrings.
    Here, we show that this self-similar structure persists in phase space.
    Moreover, near the photon shell of bound photon orbits, dynamics are controlled by a Lyapunov exponent $\gamma$, whose role we highlight by computing the first-return map for light rays close to an unstably bound orbit.
    Despite an exponential $e^\gamma$ sensitivity to initial conditions, nearly bound rays do not exhibit chaotic behavior.
    However, as the background spacetime is deformed away from the Kerr geometry, chaos sets in, with its first onset most visible near strongly resonant  bound orbits in the photon shell.
    We display two animations: one illustrating the emergence of chaos near the photon shell, which results in a fractal phase-space structure, and another exhibiting how this chaotic, fractal, self-similar structure is encoded in the first-return map.
\end{abstract}

\vspace{1em}
\tableofcontents

\parskip=1em

\section{Introduction}

Over a century ago, Albert Einstein realized that his newly formulated theory of general relativity predicted that starlight would bend slightly---by only five ten-thousandths of a degree---as it passed near the Sun \cite{Einstein:1911, Einstein:1916}.
Near a black hole, by contrast, light can be deflected not just by a few ten-thousandths of a degree but by half an orbit, a full orbit, or even multiple orbits.
Indeed, the first image of a black hole---Messier~87*---released by the Event Horizon Telescope (EHT) in 2019 revealed a now-famous ring of light around a dark ``black hole shadow'' \cite{EHT2019_I, EHT2019_IV, EHT2019_V, EHT2019_VI}; some of that light likely originated from behind the black hole, relative to Earth's line of sight.

The first nontrivial exact solution of Einstein's field equations was communicated in December 1915 by his friend, the polymath astronomer Karl Schwarzschild, from the bloody Russian front, where he was serving \cite{Schwarzschild:1916:English}.
Einstein was delighted, having previously had to rely on approximations to derive observable consequences of his theory.
Nearly half a century later, in 1963, Roy Kerr discovered another exact solution that now carries his name and which, remarkably, describes astrophysical (rotating) black holes \cite{Kerr:1963}.

Near the horizon of a Kerr black hole, where light trajectories are strongly warped, lies a central structure: the photon shell, a family of unstable spherical photon orbits that forms the separatrix between photon capture and escape \cite{Bardeen:1972, Teo:2003}.
Its imprint on a distant observer's screen is the ``photon ring'': a nested sequence of lensed subrings that together produces a ring-shaped brightness enhancement in black hole images \cite{Gralla:2019, Johnson:2020}.
These subrings arise from photons that linger in the photon shell, complete multiple orbits, and then escape toward our telescopes.

Astronomers are particularly excited about the photon ring because it offers a direct probe of black hole geometry \cite{Johnson:2020}, especially black hole spin, a key ingredient in models of galaxy-spanning relativistic jets that clear space of dust and carve out stellar voids \cite{Blandford:1977}.
A proposed space mission already aims to expand the EHT's virtual telescope to roughly three Earth diameters in order to resolve the photon ring directly \cite{BHEX:2024a, BHEX2024b}.

The photon ring is self-similar, decomposing into a sequence of successively demagnified, rotated, and time-delayed images of the source emission \cite{Gralla:2020b}.
In this work, we visualize this structure in the reduced phase space of Kerr null geodesics by numerically evolving light trajectories launched from an equatorial Poincaré section near the separatrix.
Our phase-space visualizations make the photon ring self-similarity manifest.

For a static (non-spinning) black hole, the photon shell collapses to a perfectly thin photon sphere, and all light trajectories within it are circular and identical up to rotation.
Seen from Earth, this structure would appear as a thin, bright, perfectly symmetric ring, far narrower than the EHT image of M87*.
If the black hole spins, dragging spacetime with it, the picture becomes richer and more complex.
For a spinning black hole---as most astrophysical black holes are---the photon sphere expands into a genuine shell that is thinnest near the poles and thickest near the equator (see Fig.~2 of \cite{Johnson:2020}).
Photons co-rotating with the black hole can approach closer to the horizon, whereas counter-rotating photons must orbit at larger radii to avoid plunging inward.

Photons that appear inside the photon ring follow trajectories that, when traced backward into the geometry, fall through the horizon.
Photons that appear outside the ring follow trajectories that approach the black hole, execute several orbits, and then escape.
This sharp sensitivity to initial conditions provides a necessary---but not sufficient---ingredient of chaotic dynamics.
Often called the ``butterfly effect,'' it captures the idea that arbitrarily small differences in initial conditions can lead to dramatically different outcomes.

\begin{figure}[t]
    \centering
    \includegraphics[width=0.8\textwidth]{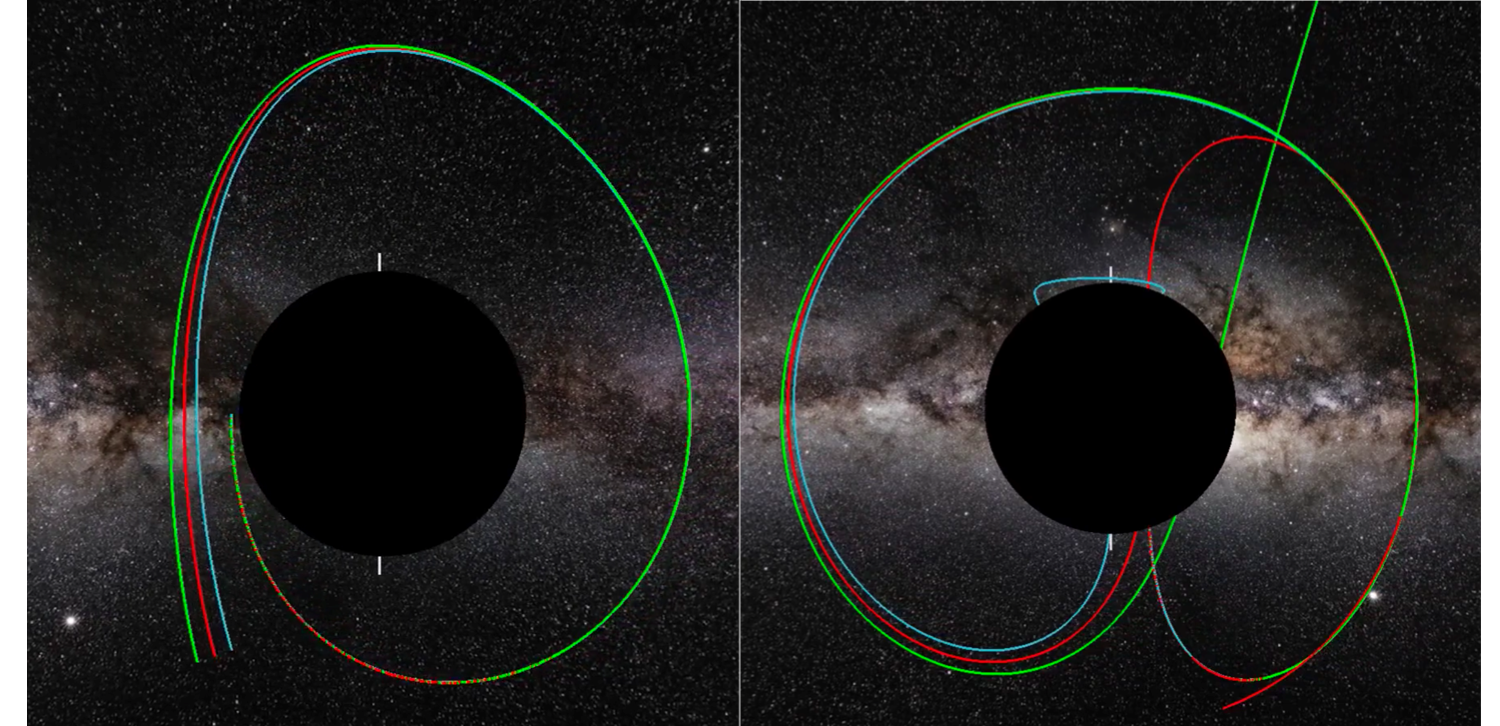}
    \caption{The fates of three nearby photons, separated by small differences in initial radius.}
    \label{fig:UltimateFates}
\end{figure}

To illustrate this behavior, Fig.~\ref{fig:UltimateFates} from \cite{Galison:2024} shows three photons launched near the horizon of a spinning black hole with identical initial momenta and nearly identical initial positions.
Initially indistinguishable, their trajectories ultimately diverge: the blue ray plunges into the black hole, the red ray asymptotes to a critical orbit, and the green ray escapes, never to return.

This sensitivity is governed by a Lyapunov exponent $\gamma$, which quantifies the rate of separation of nearby trajectories.
Introduced by the Russian mathematician Aleksandr Lyapunov to characterize instability in dynamical systems \cite{Lyapunov:1992}, this concept applies directly to the photon ring \cite{Cardoso:2009}.
In phase space, the photon ring contains a hyperbolic fixed point of the return map---the map that takes initial conditions on our equatorial Poincaré section and evolves them until they complete one orbit and return to the section.
This fixed point corresponds to an unstable spherical orbit within the photon shell and projects to a point on the photon ring.
In a sufficiently small neighborhood of this fixed point, the dynamics are governed by the linearized Hamilton equations, or equivalently, the differential of the return map.
Diagonalizing this differential yields eigenvalues $e^{\pm 2\gamma}$, where $\gamma$ is the Lyapunov exponent associated with the spherical orbit.
In this work, we numerically compute the return map, its differential, and its eigendecomposition, and use them to visualize the resulting sensitivity and self-similarity near the fixed point.

Another remarkable feature of the Kerr solution is that geodesic motion admits four constants of motion: the total energy, the axial angular momentum, the rest mass, and a fourth quantity discovered by Brandon Carter, now called the Carter constant \cite{Carter:1968}.
For the Kerr spacetime, the Carter constant completes the set of integrals of motion, rendering geodesic dynamics integrable and therefore non-chaotic.
Thus, although Kerr geodesics exhibit exponential sensitivity---a necessary ingredient for chaos---that sensitivity alone is not sufficient to produce chaotic behavior.

Now suppose that a Kerr black hole is dynamically perturbed by infalling matter or by a compact companion, or that its geometry is otherwise deformed---for example, by a stationary accretion disk, by a magnetic field, or by beyond-Einstein corrections.
Then suddently, integrability (and with it predictability) is broken.
Some or all constants of motion may be lost, the equations of motion cease to be separable, and the resulting orbits become substantially more complex.
In the generic case, this loss of integrability gives rise to the fractal, self-similar structures characteristic of chaotic dynamics, familiar from far less exotic systems.
These chaotic features have been studied in several black hole contexts: for a stationary non-Kerr spacetime~\cite{Kostaros:2022}; for a static two-black-hole toy spacetime~\cite{Shipley:2016omi}; and earlier in a fully dynamical binary black hole system~\cite{Bohn:2014xxa}.

In this work, we consider a class of continuous deformations of the Kerr spacetime and present an animation illustrating the onset of chaos in phase space.
Kolmogorov--Arnold--Moser (KAM) theory describes how this transition occurs \cite{Kolmogorov:1954, Arnold:1963, Moser:1962, Lichtenberg:1992}.
According to KAM theory, the first signatures of chaos appear near geodesic orbits with strongly resonant frequencies---those that close after only a few polar oscillations, rather than ergodically filling their shell.
For non-resonant frequencies, a finite deformation is required before chaotic behavior becomes visible.
We parameterize our deformations of the Kerr geometry by a parameter $\epsilon\in[0,1]$.
Focusing on a bound orbit that is not strongly resonant, we find that chaotic fractal structure becomes apparent only for $\epsilon\gtrsim0.99$ in our visualizations.

Chaos is also visible in \emph{escape-basin} plots.
The famous Mandelbrot fractal is, in essence, an escape-basin plot of a discrete dynamical system.
In the Mandelbrot set, each pixel is colored according to whether the corresponding point in the complex plane escapes to infinity under repeated iteration of the map $z\mapsto z^2+c$, or remains confined to a finite region.
This is analogous to a Poincaré first-return map for photons orbiting our deformed Kerr black hole.
Famously, the fate of each point---and thus each pixel color---depends sensitively on location, producing a set with a self-similar fractal boundary.

The same phenomenon appears in the escape basins of a deformed Kerr photon ring.
In this setting, there are two basins: regions of phase space from which photons plunge into the black hole and regions from which they escape to infinity.
Along the boundary between these outcomes, new basins emerge, and those basins themselves contain yet smaller basins with their own intricate boundaries.
In the chaotic regime, the smooth Kerr separatrix is replaced by a fractal basin boundary with fine-scale interleaving of capture and escape trajectories \cite{10.1098/rspa.1991.0160, Dettmann:1994dj}.

In the deformed spacetime, we again compute the differential of the return map about a fixed point and use it to analyze local dynamics.
Remarkably, the fine structure exhibits a self-similarity encoded in the action of the return map itself.
By repeatedly applying this differential to the escape basin, we show how the basin locally maps into itself under the flow, revealing the geometric mechanism underlying the fractal structure.
We present an animation of this action, demonstrating that phase-space dynamics in deformed Kerr exhibit self-similarity, exponential sensitivity to initial conditions, and genuine chaos.

\section{Null Geodesic Phase Space in Stationarity and Axisymmetry}
\label{sec:NullGeodesics}

In this section, we develop the framework used to study the phase space of null geodesics in a generic stationary, axisymmetric spacetime.
This class includes both Kerr spacetime and the deformations considered herein.

\subsection{Equations of Motion}

In a 4-dimensional spacetime, geodesic motion is often formulated in an extended 8-dimensional phase space with coordinates $(x^\mu, p_\mu) = (t, r, \theta, \phi, p_t, p_r, p_\theta, p_\phi)$ and Hamiltonian
\begin{align}
    \mathcal{H} = \frac{1}{2} g^{\mu \nu} p_\mu p_\nu.
\end{align}
Null geodesics satisfy $p^2 = 0$, so one imposes the Hamiltonian constraint $\mathcal{H} = 0$.

The line element of a stationary, axisymmetric spacetime can be written as
\begin{gather}
    \label{eq:StataxiSpacetime}
    ds^2 = g_{tt}\ed t^2 + g_{rr}\ed r^2 + g_{\theta\theta}\ed\theta^2 + 2 g_{t\phi}\ed t \ed\phi + g_{\phi\phi}\ed\phi^2,
\end{gather}
where the metric components $g_{\mu\nu}$ generally depend on $r$ and $\theta$.
This spacetime admits two Killing vectors, $\pd_t$ and $\pd_\phi$, associated with time translations and rotations about the symmetry axis, respectively.
These symmetries imply two conserved quantities along geodesics: the energy $E$ and the azimuthal angular momentum $L$.
For an affinely parametrized geodesic with four-momentum $p^\mu$, these quantities are respectively defined as
\begin{subequations}
\label{eq:EandL}
\begin{align}
    E &= -p_\mu \pd_t^\mu = -p_t,\\
    L &= p_\mu \pd_\phi^\mu = p_\phi.
\end{align}
\end{subequations}
These conserved quantities allow us to reduce the extended 8-dimensional phase space to a 4-dimensional phase space with coordinates $(r, \theta, p_r, p_\theta)$, which we denote by $\mathcal{M}$: substituting Eqs.~\eqref{eq:EandL} into the Hamiltonian eliminates $p_t$ and $p_\phi$, yielding
\begin{subequations}
\label{eq:Hamiltonian}
\begin{align}
    \mathcal{H} &= \frac{1}{2} \pa{g^{rr} p_r^2 + g^{\theta \theta} p_\theta^2 + E^2 g^{tt} + L^2 g^{\phi \phi} - 2 E L g^{t \phi}} \\
    &= \frac{1}{2} \pa{g^{rr} p_r^2 + g^{\theta \theta} p_\theta^2} + V(r, \theta),
\end{align}
\end{subequations}
where we have defined the effective potential
\begin{align}
    V \equiv \frac{1}{2} \pa{E^2 g^{tt} + L^2 g^{\phi \phi} - 2 E L g^{t \phi}}.
\end{align}
This is of the general form
\begin{align}
    \label{eq:ReducedHamiltonian}
    \mathcal{H} = \frac{1}{2} M^{ab}(x) p_a p_b + V(x),
\end{align}
provided we also define a symmetric (inverse) mass matrix
\begin{align}
    M^{ab} \equiv g^{ab} \qquad (a,b \in \{r, \theta\}).
\end{align}
Throughout this paper, Greek indices denote spacetime coordinates.
Latin indices $a,b,c,d$ label the configuration-space poloidal coordinates $\{r,\theta\}$, while Latin indices $i,j,k,\ell$ run over coordinates on the reduced phase space $(r, \theta, p_r, p_\theta)$.
Hamilton's equations are then
\begin{subequations}
\begin{align}
    \dot{x}^a &= \frac{\pd \mathcal{H}}{\pd p_a} = M^{ab} p_b, \\
    \dot{p}_a &= - \frac{\pd \mathcal{H}}{\pd x^a} = - \frac{1}{2} p_b p_c \pd_a M^{bc} - \pd_a V,
\end{align}
\end{subequations}
where over-dots denote differentiation with respect to an affine parameter.
In particular, we find
\begin{subequations}
\label{eq:Hamilton}
\begin{align}
    \label{eq:r-Hamilton}
    \dot{r} &= g^{rr} p_r, \\
    \label{eq:theta-Hamilton}
    \dot{\theta} &= g^{\theta \theta} p_\theta, \\
    \label{eq:pr-Hamilton}
    \dot{p}_r &= - \frac{1}{2} \pa{p_r^2 \pd_r g^{rr} + p_\theta^2 \pd_r g^{\theta \theta}} - \pd_r V, \\
    \label{eq:ptheta-Hamilton}
    \dot{p}_\theta &= - \frac{1}{2} \pa{p_r^2 \pd_\theta g^{rr} + p_\theta^2 \pd_\theta g^{\theta \theta}} - \pd_\theta V. 
\end{align}
\end{subequations}
Solving for $p_r$ (or $p_\theta$) in Eq.~\eqref{eq:r-Hamilton} [or~\eqref{eq:theta-Hamilton}] and substituting into Eq.~\eqref{eq:pr-Hamilton} [or~\eqref{eq:ptheta-Hamilton}] yields the equivalent second-order form of the geodesic equations,
\begin{subequations}
\begin{align}
    2 g_{rr}\ddot{r} + 2\dot{r}\dot{\theta}\pd_\theta g_{r r} 
    + \dot{r}^2\pd_r g_{rr} - \dot{t}^2\pd_r g_{tt} 
    - 2\dot{t}\dot{\phi}\pd_r g_{t\phi} 
    - \dot{\theta}^2\pd_r g_{\theta \theta} 
    - \dot{\phi}^2\pd_r g_{\phi \phi}&=0,\\
    2g_{\theta\theta} \ddot{\theta} 
    + 2\dot{r}\dot{\theta}\pd_r g_{\theta \theta} 
    - \dot{r}^2\pd_\theta g_{rr} 
    - \dot{t}^2\pd_\theta g_{t t} 
    - 2\dot{t}\dot{\phi} \pd_\theta g_{t\phi} 
    + \dot{\theta}^2\pd_\theta g_{\theta\theta} 
    - \dot{\phi}^2\pd_\theta g_{\phi \phi} &= 0.
\end{align}
\end{subequations}

As we will review in Sec.~\ref{sec:Kerr}, the Kerr spacetime possesses an additional hidden symmetry which yields an extra conserved quantity for geodesic motion: the Carter constant.
This fourth integral of motion Poisson-commutes with the conserved quantities associated with stationarity and axisymmetry, rendering the system Liouville integrable and precluding Hamiltonian chaos.
The Carter constant also appears as the separation constant in the Hamilton--Jacobi equation, allowing the $r$ and $\theta$ motions to decouple after introducing Mino time.

For generic stationary and axisymmetric spacetimes (such as the Hartle--Thorne deformations considered below), no additional hidden symmetry is present, so there is no conserved quantity beyond energy and angular momentum.
The system is therefore generically non-integrable, and chaotic behavior may arise.
At the level of the equations of motion, the $r$ and $\theta$ dynamics cannot be separated.

\subsection{Poincaré Sections, First-Return Map, and Linearized Dynamics}

A Poincaré section is a lower-dimensional surface in the phase space of a dynamical system.
In the reduced 4-dimensional phase space $\mathcal{M}$ with coordinates $(r, \theta, p_r, p_\theta)$, null geodesic trajectories lie on the codimension-1 (i.e., 3-dimensional) constraint surface $\mathcal{E} = \{\mathcal{H} = 0\}$.
We consider the 2-dimensional Poincaré section $\mathcal{S}$ formed by the intersection of this constant-energy surface with the set of downgoing ($p_\theta>0$) orbits through the equatorial plane $\Sigma = \{\theta = \pi/2\}$, that is, $\mathcal{S} = \mathcal{E} \cap \Sigma \subset \mathcal{M}$.
The coordinates $(r, p_r)$ parameterize points on $\mathcal{S}$ because $\theta = \pi/2$ and solving $\mathcal{H} = 0$ determines the downgoing root $p_\theta>0$ once $(r, p_r, \theta)$ are specified.

Photon trajectories in phase space with initial conditions on $\mathcal{S}$ have three qualitative outcomes.
They may (1) escape to infinity before returning to $\mathcal{S}$, (2) cross the event horizon before returning to $\mathcal{S}$, or (3) return to $\mathcal{S}$ one or more times---either infinitely many times (periodic orbits, bound motion) or a finite nonzero number of times before eventual escape or capture.

Let $\xi^i = (r, \theta, p_r, p_\theta)$ denote the phase-space coordinates.
Given a Hamiltonian $\mathcal{H}$, the associated Hamiltonian vector field $X_\mathcal{H}$ has components
\begin{align}
    \label{eq:HamiltonianField}
    X^i_\mathcal{H} = \{ \xi^i, \mathcal{H} \} = \frac{d}{dt} \xi^i,
\end{align}
where $\lambda$ is the time parameter along the trajectory associated with $\mathcal{H}$.\footnote{For the Hamiltonian \eqref{eq:Hamiltonian}, this time parameter is not Boyer-Lindquist time, but affine time.
For a generic Hamiltonian, $t$ need not be affine.
For instance, the Hamiltonian $\mathcal{H}_M=\Sigma \mathcal{H}$ time-evolves according to Mino time $\tau$ as in Eqs.~\eqref{eq:MinoTime}.}
If the $\xi^i$ are canonical coordinates on $\mathcal{M}$, these are precisely Hamilton's equations (i.e., the components of $X^i_{\mathcal{H}}$ are given by the right-hand sides of Eqs.~\eqref{eq:Hamilton}).
The integral curves of $X^i_{\mathcal{H}}$ are the phase-space trajectories generated by $\mathcal{H}$.

\begin{figure}[t]
    \centering
    \resizebox{0.8\textwidth}{!}{\definecolor{sectioncol}{RGB}{85,85,85}%
\definecolor{trajcol}{RGB}{0,169,0}%
\definecolor{devveccol}{RGB}{0,0,255}%
\definecolor{tangellipsecol}{RGB}{255,0,0}%
\tikzset{
  arrowmark/.style={
    postaction={decorate},
    decoration={
      markings,
      mark=#1 with {\arrow{>}}}},
  thinner/.style={line width=0.106},
  mydashed/.style={dash pattern=on 0.21pt off 0.21pt}
}
\def \globalscale {1.250000}%
\begin{tikzpicture}[y=0.80pt, x=0.80pt,
  yscale=-\globalscale, xscale=\globalscale,
  >={Stealth[length=1.pt]},
  line width=0.212pt]

  \path[draw=tangellipsecol] (23.40,11.13) ellipse (0.12cm and 0.06cm);

  \path[draw=sectioncol]
  (23.53,3.65) .. controls (19.79,3.97) and (19.02,4.97) .. (14.68,5.18) .. controls (13.64,8.56) and (8.94,14.18) .. (5.45,20.58) .. controls (10.08,18.90) and (35.27,19.31) .. (43.64,19.64) .. controls (46.09,12.44) and (46.83,8.55) .. (49.41,3.68) .. controls (42.41,4.88) and (28.46,3.26) .. (24.91,3.53);

  \coordinate (initp) at (23.74,10.93) ;
  \coordinate (initptip) at (27.64,10.86) ;

  \path[->, draw=devveccol] (initp) -- (initptip);

  \filldraw[trajcol] (initp) circle (0.25pt);
  \path[arrowmark={between positions .2 and .95 step 1cm}, draw=trajcol]
  (initp) .. controls (25.97,3.42) and (24.33,0.08) .. (17.01,0.19) .. controls (9.68,0.31) and (1.63,7.00) .. (0.84,12.67) .. controls (0.04,18.35) and (1.07,20.80) .. (4.77,24.00) .. controls (8.47,27.19) and (16.56,29.19) .. (23.52,27.98) .. controls (28.55,26.79) and (32.66,24.31) .. (34.45,19.86);

  \coordinate (finalp) at (35.92,12.29) ;

  \path[draw=trajcol, mydashed]
  (34.8,19.04) .. controls (35.5,17.82) and (finalp) .. (finalp);

  \begin{scope}[/pgf/fpu/install only={reciprocal}]
    \path[arrowmark={at position 0.65},
    draw=trajcol, thinner]
    (initptip) .. controls (initptip) and (27.56,9.72) .. (27.56,9.18) .. controls (27.56,8.43) and (27.51,7.79) .. (27.51,7.79);
  \end{scope}

  \coordinate (finalptip) at (41.65,12.00) ;

  \path[->, draw=sectioncol] (finalp) -- (finalptip);

  \coordinate (imagetangref) at (37.9,14.);
  \newcommand{\imtanAngA}{85}
  \newcommand{\imtanAngB}{260}
  \newcommand{\imtanRadA}{5.76}
  \newcommand{\imtanRadB}{2.32}

  \path[cm={{0.84,-0.54,0.44,0.90,(0.0,0.0)}},draw=tangellipsecol,mydashed]
  (imagetangref)arc(\imtanAngA:\imtanAngB:{\imtanRadA} and {\imtanRadB});

  \path[draw=sectioncol] (36.17,12.47) ellipse (0.16cm and 0.05cm);

  \begin{scope}[/pgf/fpu/install only={reciprocal}]
    \path[arrowmark={at position .8},
    draw=trajcol, thinner]
    (finalptip) .. controls (41.50,10.96) and (41.13,9.46) .. (40.80,8.31) .. controls (40.38,6.76) and (39.88,6.00) .. (39.88,6.00);
  \end{scope}

  \path[cm={{0.84,-0.54,0.44,0.90,(0.0,0.0)}},draw=tangellipsecol]
  (imagetangref)arc(\imtanAngB:\imtanAngA:{-\imtanRadA} and {-\imtanRadB});

  \path[->, draw=devveccol] (finalp) -- (41.,9.5);

  \begin{scope}[/pgf/fpu/install only={reciprocal}]
    \filldraw[trajcol] (finalp) circle (0.25pt);
    \path[arrowmark={at position .7}, draw=trajcol]
    (finalp) .. controls (35.83,11.32) and (35.99,10.71) .. (35.40,8.89) .. controls (35.06,7.85) and (34.45,6.85) .. (33.72,5.74);
  \end{scope}

\end{tikzpicture}}
    \caption{Schematic illustration of the first-return map on the Poincaré section $\mathcal{S}$.
    A base point $p\in\mathcal{S}$ (green dot, left) returns to $\mathcal{S}$ after an affine time $T(p)$ under the Hamiltonian flow (large green orbit).
    A small circular bundle of nearby initial conditions $\mathcal{C}$ (solid red circle) is co-evolved for the same amount of time.
    At time $T(p)$, some points have not yet returned to $\mathcal{S}$ (dashed red segment), while others have already crossed it.
    Projecting the evolved bundle back onto $\mathcal{S}$ yields the image ellipse $\tilde{\mathcal{C}}$ (gray ellipse, right), which represents the action of the differential of the return map $dF_p$ on infinitesimal deviations (blue vectors).
    Because the flow is symplectic, the first-return map is area-preserving.}
    \label{fig:PoincaréSection}
\end{figure}
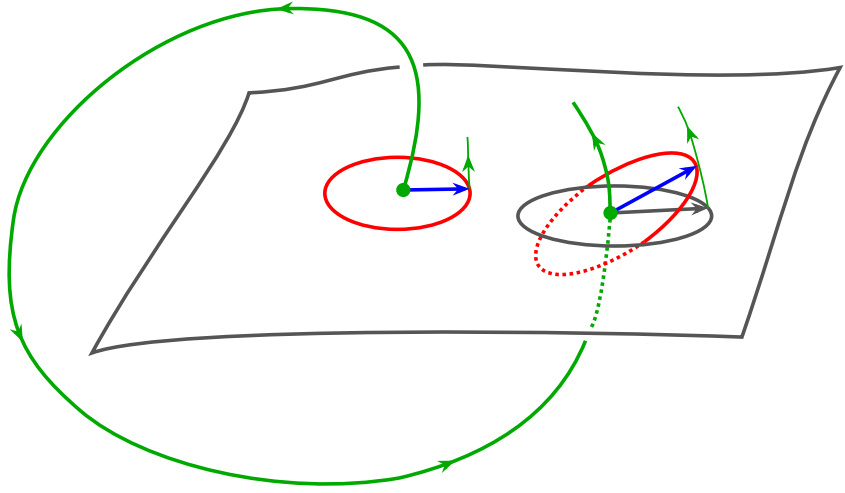

Photon trajectories launched from $\mathcal{S}$ that do return to $\mathcal{S}$ (that is, which are in class 3 above) do so after a trajectory-dependent affine time.
If we evolve a small circle of initial conditions on $\mathcal{S}$---representing a narrow bundle of light rays---together with a base point $p \in \mathcal{S}$, then after the return time $T(p)$ of the base point, neighboring trajectories will generally not lie on $\mathcal{S}$: some will not yet have reached the section, while others will already have crossed it.
This is illustrated in Fig.~\ref{fig:PoincaréSection}: the initial bundle is the solid red circle on the left, centered on the base point $p$ (green), while its co-evolved image at time $T(p)$ is shown on the right as a red ellipse that does not fully lie within $\mathcal{S}$: some of its (dashed red) points have not yet reached $\mathcal{S}$, while others (solid red arc) have already evolved off of it.
Projecting that image back onto $\mathcal{S}$ produces the (black) ellipse that represents the linearized action of the return map near $p$.

These notions can be defined rigorously as follows.
Let $\Phi_t: \mathcal{M} \to \mathcal{M}$ denote the time-evolution map that evolves a point $p \in \mathcal{M}$ by an amount $t$ along the integral curve of $X_\mathcal{H}$ through $p$.
Let $\tilde{\mathcal{S}} \subseteq \mathcal{S}$ denote the subset of points whose trajectories return to $\mathcal{S}$ at least once under this flow.
For each $p \in \tilde{\mathcal{S}}$, let $T(p)$ denote the first-return time, so that $\Phi_{T(p)}(p) \in \mathcal{S}$.
The first-return map $F: \tilde{\mathcal{S}} \to \mathcal{S}$ is then defined by
\begin{align}
    F(p) = \Phi_{T(p)}(p).
\end{align}
To describe the evolution of nearby initial conditions, we linearize Eq.~\eqref{eq:HamiltonianField} about a trajectory.
Infinitesimal deviations $\delta\xi^i$ (the blue vectors in Fig.~\ref{fig:PoincaréSection}) then satisfy the linear system
\begin{align}
    \frac{d}{d\lambda} \delta \xi^i = \pa{\pd_j X^{i}_\mathcal{H}} \delta \xi^j \equiv \mathcal{J}^{i}{}_j \delta \xi^j,
\end{align}
where $\mathcal J^i{}_j=\pd_j X^i_{\mathcal H}$.
For the Hamiltonian \eqref{eq:ReducedHamiltonian}, this Jacobian is straightforward to compute.
Since $X^i_{\mathcal H}=\{\xi^i,\mathcal H\}$, we have
\begin{subequations}
\begin{align}
    X^{x^{a}}_{\mathcal{H}} &= \frac{\pd \mathcal{H}}{\pd p_a} = M^{ab} p_b, \\
    X^{p_a}_{\mathcal{H}} &= - \frac{\pd \mathcal{H}}{\pd x^a} = - \frac{1}{2} p_b p_c \pd_a M^{b c} - \pd_a V.
\end{align}
\end{subequations}
Next, we compute
\begin{subequations}
\begin{align}
    \pd_{x^b} X^{x^a}_{\mathcal{H}} &= p_c \pd_b M^{ac}, \\
    \pd_{p_b} X^{x^a}_{\mathcal{H}} &= M^{ab}, \\
    \pd_{x^b} X^{p_a}_{\mathcal{H}} &= - \frac{1}{2} p_c p_d \pd_a \pd_b M^{cd} - \pd_a \pd_b V, \\
    \pd_{p_b} X^{p_a}_{\mathcal{H}} &= - p_c \pd_a M^{bc}.
\end{align}
\end{subequations}
This provides the entries of $\mathcal{J}^{i}{}_j$.
The linearized dynamics for stationary, axisymmetric spacetimes are given by
\begin{subequations}
\label{eq:LinearizedDynamics}
\begin{align}
    \dot{\delta r} &= 
        \pa{p_r \pd_r g^{rr}} \delta r 
        + \pa{p_r \pd_\theta g^{rr}} \delta \theta
        + \pa{g^{rr}} \delta p_r, \\
    \dot{\delta \theta} &= 
        \pa{p_\theta \pd_r g^{\theta \theta}} \delta r 
        + \pa{p_\theta \pd_\theta g^{\theta \theta}} \delta \theta
        + \pa{g^{\theta \theta}} \delta p_\theta, \\
    \dot{\delta p_r} &= 
        - \br{\frac{1}{2} \pa{p_r^2 \pd_r^2 g^{rr} + p_\theta^2 \pd_r^2 g^{\theta \theta}} + \pd_r^2 V} \delta r 
        - \br{\frac{1}{2} \pa{p_r^2 \pd_\theta \pd_r g^{rr} + p_\theta^2 \pd_\theta \pd_r g^{\theta \theta}} + \pd_\theta \pd_r V} \delta \theta \notag \\
        &\phantom{=}\ - \pa{p_r \pd_r g^{rr}} \delta p_r 
        - \pa{p_\theta \pd_r g^{\theta \theta}} \delta p_\theta, \\
    \dot{\delta p_\theta} &= 
        - \br{\frac{1}{2} \pa{p_r^2 \pd_r \pd_\theta g^{rr} + p_\theta^2 \pd_r \pd_\theta g^{\theta \theta}} + \pd_r \pd_\theta V} \delta r 
        - \br{\frac{1}{2} \pa{p_r^2 \pd_\theta^2 g^{rr} + p_\theta^2 \pd_\theta^2 g^{\theta \theta}} + \pd_\theta^2 V} \delta \theta \notag \\
        &\phantom{=}\ - \pa{p_r \pd_\theta g^{rr}} \delta p_r 
        - \pa{p_\theta \pd_\theta g^{\theta \theta}} \delta p_\theta.
\end{align}
\end{subequations}
Since the Jacobian $\mathcal{J}^{i}{}_j$ depends on the phase-space point, the linearized system must be evolved simultaneously with the full nonlinear equations \eqref{eq:Hamilton}.

Next, consider a small circular beam of initial conditions on $\mathcal{S}$, as in Fig.~\ref{fig:PoincaréSection},
\begin{align}
    \mathcal{C} = \{ (r^*, p_r^*) + \rho\, (\cos \varphi, \sin \varphi) : 0 \leq \varphi \leq 2 \pi \} \subset \mathcal{S},
\end{align}
centered on a point $p = (r^*, p_r^*) \in \mathcal{S}$, with $\rho>0$ sufficiently small.
For each $q = (r, p_r) \in \mathcal{C}$, one may compute $F(q)$ either by evolving the full nonlinear equations with initial condition $q$, or by evolving the deviation $\delta \xi^i = \xi_q^i - \xi_p^i$ according to Eqs.~\eqref{eq:LinearizedDynamics} for a time $T(p)$ and then setting $F(q) = \xi^i_{F(p)} + \delta \xi^i (T(p))$.\footnote{There are two technicalities that one must take into account: (1) a point in $\mathcal{C}$ specifies only $\delta r$ and $\delta p_r$, so one must additionally impose tangency to $\mathcal{S}$ to obtain $\delta \theta$ and $\delta p_\theta$, thereby embedding the deviation into the tangent space of $\mathcal{M}$; and (2) the deviation will in general pick up components that are not tangent to $\mathcal{S}$ after evolution to first return, and one must project back onto the Poincaré section before adding the deviation to $\xi^i_{F(p)}$ (the projected ellipse is the gray ellipse on the right-hand side of of Fig.~\ref{fig:PoincaréSection}).
See App.~\ref{app:TangentEvolution} and the discussion in App.~\ref{app:PoincaréMaps}.}
These two procedures are equivalent to linear order, and they both map the beam $\mathcal{C}$ to an ellipse $\tilde{\mathcal{C}}$.
Since $\mathcal{H}$ generates a symplectic flow, the induced first-return map is area-preserving, and hence $\mathcal{C}$ and $\tilde{\mathcal{C}}$ have the same area; see App.~\ref{app:PoincaréMaps} for a proof.

\section{Kerr: The Photon Ring in Phase Space}
\label{sec:Kerr}

In this section, we introduce the Kerr spacetime and apply the framework developed in Sec.~\ref{sec:NullGeodesics} to analyze the photon ring in phase space.

\subsection{The Kerr Spacetime}

The Kerr metric describes the spacetime geometry around an astrophysical rotating black hole.
In Boyer--Lindquist coordinates $(t,r,\theta,\phi)$, its line element $ds^2=g_{\mu\nu}\ed x^\mu \ed x^\nu$ is
\begin{subequations}
\label{eq:Kerr}
\begin{gather}
    ds^2=\pa{-\frac{\Delta}{\Sigma}\pa{\ed t-a\sin^2{\theta}\ed\phi}^2+\frac{\Sigma}{\Delta}\ed r^2+\Sigma\ed\theta^2+\frac{\sin^2{\theta}}{\Sigma}\br{\pa{r^2+a^2}\ed\phi-a\ed t}^2},\\
    \Delta=r^2-2Mr+a^2,\qquad
    \Sigma=r^2+a^2\cos^2{\theta}
\end{gather}
\end{subequations}
This is of the form \eqref{eq:StataxiSpacetime}, so the Kerr geometry is stationary and axisymmetric, with $\pd_t$ and $\pd_\phi$ generating time translations and azimuthal rotations, respectively.
The general framework and dynamical equations developed in Sec.~\ref{sec:NullGeodesics} therefore apply directly to the Kerr geometry.

In Kerr, it is convenient to go beyond the general reduction described in Sec.~\ref{sec:NullGeodesics} by working directly in the reduced phase space $(r, \theta, p_r, p_\theta)$ with a Hamiltonian that generates evolution with respect to Boyer--Lindquist coordinate time $t$.
This is achieved by solving the null condition $p_\mu p^\mu=0$ for $p_t$ and defining a new Hamiltonian $H(r, \theta, p_r, p_\theta) = - p_t$ (see App.~\ref{app:TimeEvolution} for a brief justification), so that $H = - p_t \equiv E$ is the conserved energy in this formulation.
One then finds \cite{Hadar:2022}
\begin{subequations}
\begin{gather}
    H(r, \theta, p_r, p_\theta) = \br{\frac{\pa{r^2+a^2}}{\Delta} - a^2\sin^2{\theta}}^{-1} \pa{\frac{2Mar}{\Delta}L+\sqrt{G}}, \\
    G = \pa{\frac{2Mar}{\Delta}L}^2 + \br{ \frac{\pa{r^2+a^2}^2}{\Delta} - a^2 \sin^2{\theta}}\br{\Delta p_r^2 + p_\theta^2 + \pa{\frac{1}{\sin^2{\theta}} - \frac{a^2}{\Delta}} L^2}.
\end{gather}
\end{subequations}
The phase-space coordinates $(r, \theta, p_r, p_\theta)$ are canonical, and the induced symplectic form is 
\begin{align}
    \omega = \ed r \wedge \ed p_r + \ed\theta \wedge \ed p_\theta,
\end{align}
with Poisson brackets given by
\begin{align}
    \{ f, g \} = \frac{\pd f}{\pd r} \frac{\pd g}{\pd p_r} 
    - \frac{\pd f}{\pd p_r} \frac{\pd g}{\pd r} 
    + \frac{\pd f}{\pd \theta} \frac{\pd g}{\pd p_\theta} 
    - \frac{\pd f}{\pd p_\theta} \frac{\pd g}{\pd \theta}.
\end{align}
As anticipated at the end of Sec.~\ref{sec:NullGeodesics}, Kerr also admits, in addition to the conserved quantities $E$ and $L$, the Carter constant
\begin{subequations}
\begin{align}
    \label{eq:r-CarterConstant}
    Q &= - \Delta p_r^2 + \frac{\br{H \pa{r^2 + a^2} - a L}^2}{\Delta} - \pa{L - a H}^2 \\
    \label{eq:theta-CarterConstant}
    &= p_\theta^2 - a^2 H^2 \cos^2{\theta} + L^2 \cot^2\theta
\end{align}
\end{subequations}
which Poisson-commutes with $E = H$ and $L$.

Inverting Eqs.~\eqref{eq:r-CarterConstant} and \eqref{eq:theta-CarterConstant} yields decoupled expressions for $p_r$ and $p_\theta$:
\begin{subequations}
\begin{align}
    p_r &= \frac{\pm \sqrt{\mathcal{R}(r)}}{\Delta},
    &\mathcal{R}(r) &= \br{H \pa{r^2 + a^2} - a L}^2 - \Delta \br{Q + (L - a H)^2},\\
    p_\theta &= \pm \sqrt{\Theta(\theta)},
    &\Theta(\theta) &= Q + L^2 + a^2 H^2 \cos^2{\theta} - \frac{L^2}{\sin^2{\theta}}.
\end{align}
\end{subequations}
Using these expressions, we recall that for an affine parametrization,
\begin{subequations}
\begin{align}
    \dot{r} &= g^{rr} p_r = \frac{\pm \sqrt{\mathcal{R}(r)}}{\Sigma} , \\
    \dot{\theta} &= g^{\theta \theta} p_\theta = \frac{\pm \sqrt{\Theta(\theta)}}{\Sigma}.
\end{align}
\end{subequations}
The Mino-time parameterization $\tau$ is defined relative to an affine parameter $\lambda$ by $d\tau = d\lambda / \Sigma$.
In terms of $\tau$, the equations simplify to a pair of decoupled quadratures,
\begin{subequations}
\label{eq:MinoTime}
\begin{align}
    \frac{dr}{d\tau} &= \pm \sqrt{\mathcal{R}(r)}, \\
    \frac{d \theta}{d \tau} &= \pm \sqrt{\Theta(\theta)}.
\end{align}
\end{subequations}
This makes the integrability of the Kerr geodesic equations manifest.

\subsection{Review of the Photon Ring}

This subsection offers a brief overview of the photon ring, drawing upon the treatments in \cite{Johnson:2020, Gralla:2020a, Gralla:2020b, Lupsasca2024}, to which we refer the reader for further details.

Bound photon orbits at fixed radius $\tilde{r}$ are obtained by requiring the radial potential $\mathcal{R}(r)$ to have a double root:
\begin{align}
    \mathcal{R}(\tilde{r}) = \mathcal{R}'(\tilde{r}) = 0.
\end{align}
Introducing the energy-rescaled quantities $\lambda = L / E$ and $\eta = Q / E^2$, this condition implies
\begin{subequations}
\label{eq:Criticality}
\begin{align}
    \label{eq:lambda-Critical}
    \tilde{\lambda} &= a + \frac{\tilde{r}}{a} \br{\tilde{r} - \frac{2 \Delta}{\tilde{r}-M}}, \\
    \tilde{\eta} &= \frac{\tilde{r}^3}{a^2} \br{ \frac{4 M \Delta}{\pa{\tilde{r}-M}^2} - \tilde{r} },
\end{align}
\end{subequations}
with $\tilde{r}$ restricted to the interval $[\tilde{r}_-, \tilde{r}_+]$, where
\begin{align}
    \tilde{r}_\pm = 2 M \br{1+\cos\pa{\frac{2}{3}\arccos\pa{\pm \frac{a}{M}}}}.
\end{align}
Throughout, we use tildes to indicate quantities evaluated on these critical bound orbits.

Bound photons with critical values $\tilde{\lambda}$ and $\tilde{\eta}$ orbiting at a radius $\tilde{r}\in[\tilde{r}_-,\tilde{r}_+]$ oscillate between polar turning points.
Defining
\begin{align}
    u_\pm = \Delta_\theta \pm \sqrt{\Delta_\theta^2 + \frac{\eta}{a^2}}, \qquad 
    \Delta_\theta = \frac{1}{2} \pa{1 - \frac{\eta + \lambda^2}{a^2}},
\end{align}
the motion is confined between
\begin{align}
    \theta_\pm = \arccos \pa{\mp \sqrt{u_+}}.
\end{align}
The set of such bound photon orbits therefore forms a spherical shell of variable thickness, known as the ``photon shell.''
In the nonrotating limit $a \to 0$, this shell degenerates into a single ``photon sphere'' at $r = 3 M$.
It is also useful to define the fractional number of half-orbits,
\begin{align}
    \label{eq:FractionalOrbits}
    n = \frac{a \sqrt{- u_-}}{2 K \pa{\frac{u_+}{u_-}}} I_r,
\end{align}
where $K(m)\equiv\int_0^{\pi/2}\pa{1-m\sin^2{\theta}}^{-1/2}\ed\theta$ is the complete elliptic integral of the first kind, while
\begin{align}
    I_r = \fint_{r_{\rm s}}^{r_{\rm o}}\frac{\ed r}{\pm_r \sqrt{\mathcal{R}(r)}} 
\end{align}
is the path integral along the photon trajectory from source radius $r_\mathrm{s}$ to observer radius $r_{\rm o}$, with $\pm_r = \mathrm{sign}\pa{p^r}$ the sign of the radial component of the four-momentum.
With this convention, a trajectory completed one half-orbit as it travels from one polar turning point to the other.

Consider a precisely critical ray with conserved quantities $\tilde{\lambda}(\tilde{r})$ and $\tilde{\eta}(\tilde{r})$.
For nearby rays satisfying $|r - \tilde{r}| \ll \tilde{r}$, the separation grows exponentially with the number of half-orbits:
\begin{align}
    \frac{r_2 - \tilde{r}}{r_1 - \tilde{r}} \approx e^{2 \gamma (n_2 - n_1)}.
\end{align}
Here, the Lyapunov exponent is
\begin{align}
    \gamma = \frac{4 \tilde{r} \sqrt{\tilde{\chi}}}{a\sqrt{- \tilde{u}_-}} \tilde{K}, \qquad
    \tilde{\chi} = 1 - \frac{M \Delta}{\tilde{r}\pa{\tilde{r}-M}^2}.
\end{align}
It governs the sensitivity of null geodesics to initial conditions in the vicinity of the photon shell.

Now consider a distant observer at inclination $\theta_{\rm o} \in [0, \pi / 2)$. The observer's image plane may be parameterized by orthogonal impact parameters $(\alpha,\beta)$, which we take to be
\begin{subequations}
\label{eq:ScreenCoordinates}
\begin{align}
    \alpha &= - \frac{\lambda}{\sin{\theta_{\rm o}}} \\
    \label{eq:beta}
    \beta &= \pm_{\rm o} \sqrt{\eta + a^2 \cos^2{\theta_{\rm o}} - \lambda^2 \cot^2{\theta_{\rm o}} },
\end{align}
\end{subequations}
where $\pm_{\mathrm{o}} = \mathrm{sign}\pa{p^\theta}$.
Together, Eqs.~\eqref{eq:Criticality} and \eqref{eq:ScreenCoordinates} show that radii $\tilde{r}$ in the photon shell map to a 1-dimensional curve on the observer's image plane, provided that $\beta$ in Eq.~\eqref{eq:beta} is real. 
This curve is the critical curve; see Fig.~1 of \cite{Gralla:2020b} for an illustration.

Photon trajectories corresponding to points on the critical curve have, by definition, the critical values $\tilde{\lambda}$ and $\tilde{\eta}$.
When evolved backward from the observer, such rays asymptote to bound photon orbits in the shell that execute $n\to\infty$ half-orbits.
In optically thin emission models, successive lensed images of the source form a nested sequence of narrow subrings on the observer's screen that accumulate on the critical curve, with each successive image exponentially demagnified and exponentially pushed toward this curve.
Each subring may be labeled by the number of half-orbits executed by the corresponding photon trajectories: the $n=0$ ring is the direct image of the source, the $n=1$ ring consists of photons that execute one half-orbit, and so on.
Together, the full collection of these demagnified subring images produces a characteristic brightness, which we refer to as the ``photon ring'' \cite{Johnson:2020}.

\subsection{The Photon Ring in Phase Space}
\label{subsec:PhaseSpaceRing}

\begin{figure}[t]
    \centering
    \includegraphics[width=\textwidth]{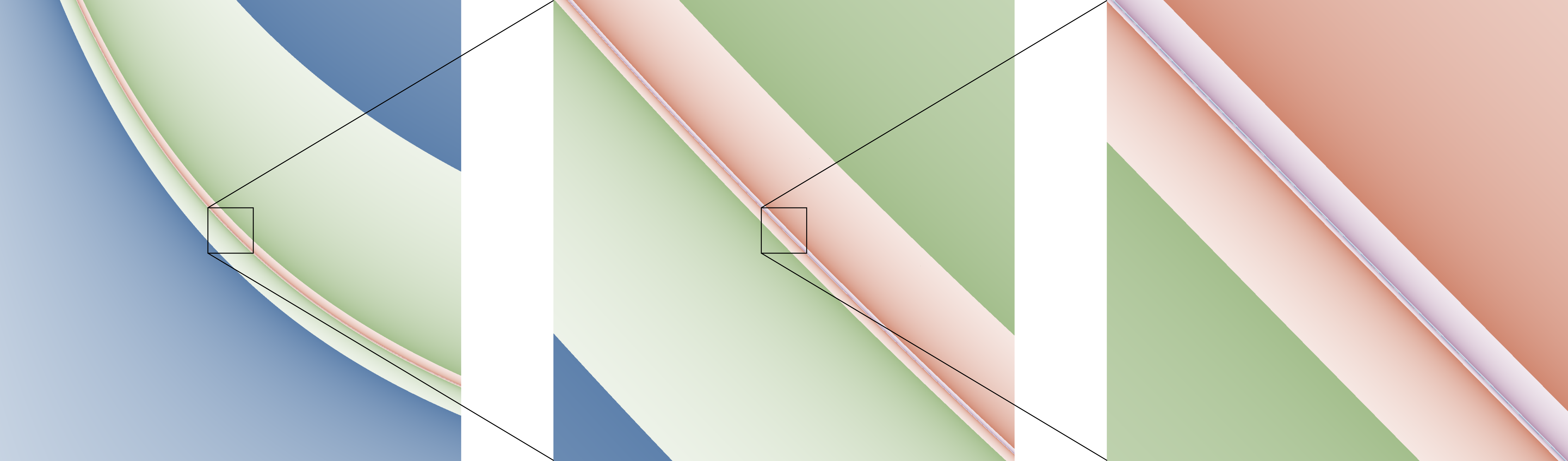}
    \caption{Triptych of the equatorial Poincar\'e section $\mathcal{S}$ for Kerr null geodesics near the unstable spherical photon orbit $(r, p_r) = (r_c, 0)$, illustrating the photon ring in phase space.
    The boxes mark the region enlarged in the next panel; from left to right, each panel zooms in by a factor of $10$.
    The diagonal boundary is the separatrix associated with the hyperbolic fixed point.
    The blue, green, red, and purple bands correspond to the $n=0$, $1$, $2$, and $3$ photon rings, respectively, while the saturation encodes the fractional part of $n$.
    The $n=0$ and $n=1$ rings do not yet lie in the universal large-$n$ regime, which begins around $n\gtrsim2$.
    In that regime, the rings exhibit a simple self-similar structure governed by the Lyapunov exponent of the bound orbit.}
    \label{fig:PhaseSpacePhotonRing}
\end{figure}

To visualize the Kerr photon shell and photon ring in phase space, we numerically evolve a region of initial conditions on $\mathcal{S}$ in the $(r, p_r)$ plane using the dynamical equations \eqref{eq:Hamilton} with the metric components determined by Eq.~\eqref{eq:Kerr}. 
Throughout, we fix $E = 1$ and $L = 4.30405 M$, and set the spin to $a = 0.327352 M$.\footnote{We choose these numerical values to match those used by Kostaros and Pappas \cite{Kostaros:2022}, so that upon deforming the spacetime into the chaotic regime, we recover visually rich structures similar to those shown in their Fig.~3.}
Our plots are centered at $p_r = 0$ and at the critical radius $r_c = 2.610235 M$, obtained from Eq.~\eqref{eq:lambda-Critical} for the chosen values of $E$ and $L$.
Each trajectory is evolved to a large affine time in order to capture its asymptotic behavior.
For every trajectory, we compute the number of half-orbits using Eq.~\eqref{eq:FractionalOrbits}.
Distinct colors indicate the integer part of $n$, while the saturation encodes its fractional part.
In our plots, the vertical axis is $\hat{p}_r \equiv p_r / r_c$, that is, $p_r$ measured in units of $r_c$.
We present a triptych whose panels have one-sided widths $(\Delta r, \Delta \hat{p}_r) = (3 \times 10^{-1} M, 3 \times 10^{-1} M)$, $(3 \times 10^{-2} M, 3 \times 10^{-2} M)$, and $(3 \times 10^{-3} M, 3 \times 10^{-3} M)$, so that each successive panel zooms in by a factor of $10$.
The result is shown in Fig.~\ref{fig:PhaseSpacePhotonRing}.

In Fig.~\ref{fig:PhaseSpacePhotonRing}, we display part of the phase-space structure of the Kerr photon ring.
Each orbital radius $\tilde{r}$ in the photon shell---equivalently, each point on the critical curve in the image plane of a distant observer---has an associated phase-space diagram; here, we show the one corresponding to our chosen value of $r_c$.
The central point marks the intersection of the bound orbit with the equatorial plane.
The approximately diagonal curve running across each panel is the phase-space separatrix representing the photon trajectories that asymptote to the bound orbit at radius $r_c$.
Points on one side of this separatrix (to its upper right) correspond to nearly bound photons that eventually escape after passing close to the photon shell, while points on the other side (to its lower left) correspond to photons that are ultimately captured by the black hole.

The blue pixels correspond to photons with $n=0$: they do not complete even a half-orbit around the black hole and therefore carry the direct image of the source to a distant observer.
The green pixels correspond to $n=1$ photons, which execute one half-orbit and produce the first lensed image, forming a narrow and bright $n=1$ photon ring.
The red and purple pixels correspond to $n=2$ and $n=3$ photons, which execute two and three half-orbits, respectively.
These trajectories produce exponentially narrower rings on the observer's screen, each representing a full demagnified image of the source around the black hole.
As one zooms in, this subring structure becomes manifestly self-similar, particularly in the transition from the second to the third panel.
In the asymptotic regime of large orbit number, $n\to\infty$, successive rings are related by simple relations.
This universal behavior is already apparent for $n\gtrsim2$, whereas the $n=0$ and $n=1$ rings still show visible departures from the asymptotic pattern, as seen in the transition from the first to the second panel.

\begin{figure}[t]
    \centering
    \includegraphics[width=\textwidth]{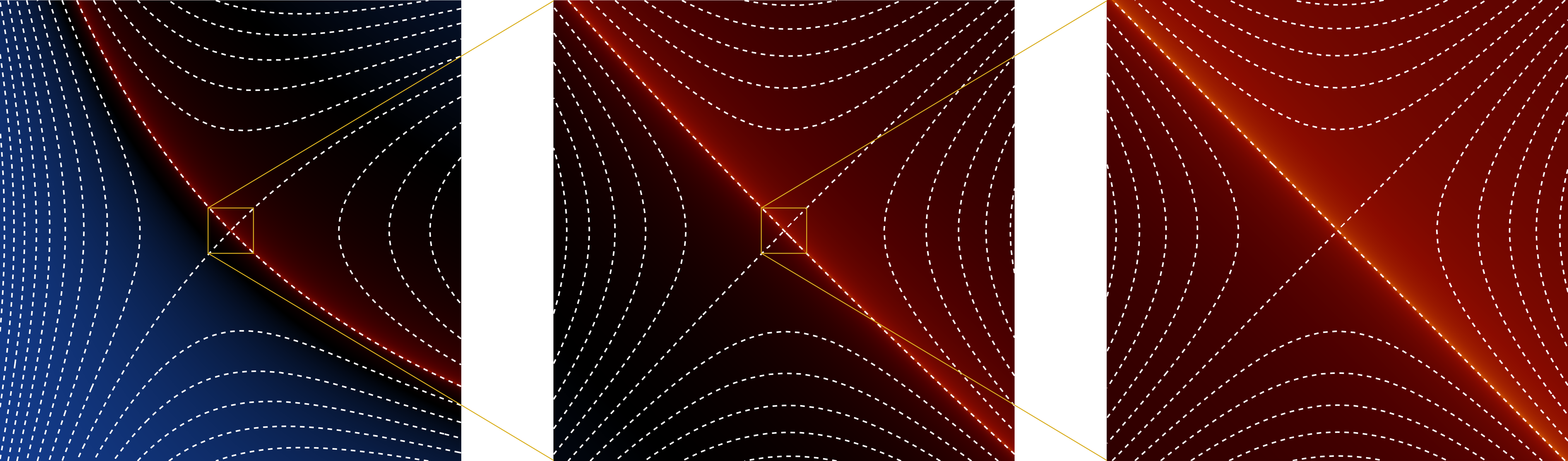}
    \caption{Triptych of the equatorial Poincar\'e section $\mathcal{S}$ for Kerr null geodesics near the unstable spherical photon orbit $(r, p_r) = (r_c, 0)$.
    Pixels are colored by the half-orbit count $n$, with blue, black, red, and orange corresponding to $n=0$, $1$, $2$, and $3$, respectively.
    The dashed curves are contours of the energy-rescaled Carter constant $\eta = Q/E^2$, which is preserved under the first-return map.
    At sufficiently small scales, the phase-space structure approaches the scale-invariant normal form of a hyperbolic fixed point, making the local self-similarity manifest.}
    \label{fig:KerrReturnMap}
\end{figure}

Next, we present a second triptych in which we overlay contours of the energy-rescaled Carter constant, $\eta = Q/E^2$ from Eq.~\eqref{eq:r-CarterConstant}, on the same phase-space region.
In this figure, the number of half-orbits $n$ is shown using a continuous color scale rather than the discrete saturation scheme used in Fig.~\ref{fig:PhaseSpacePhotonRing}.
The result is displayed in Fig.~\ref{fig:KerrReturnMap}.

Because the Carter constant is conserved, any photon that returns to the equatorial plane after completing a full polar oscillation must reappear at a point lying on the same dashed contour on which it began.
The existence of these continuous invariant curves is a hallmark of the integrability of Kerr geodesic motion.
The bound photon orbit at the center of the plots is unstable, so the origin is a hyperbolic fixed point of the return map $F$.
In the left panel, which shows the widest view, the phase-space structure retains details specific to the Kerr dynamics.
After one zoom, in the middle panel, it takes on the familiar form of the phase portrait of an inverted harmonic oscillator, the normal form near a hyperbolic fixed point.
That normal form is scale-invariant, and accordingly the third panel is qualitatively unchanged under further zoom.

\begin{figure}[t]
    \centering
    \includegraphics[width=\textwidth]{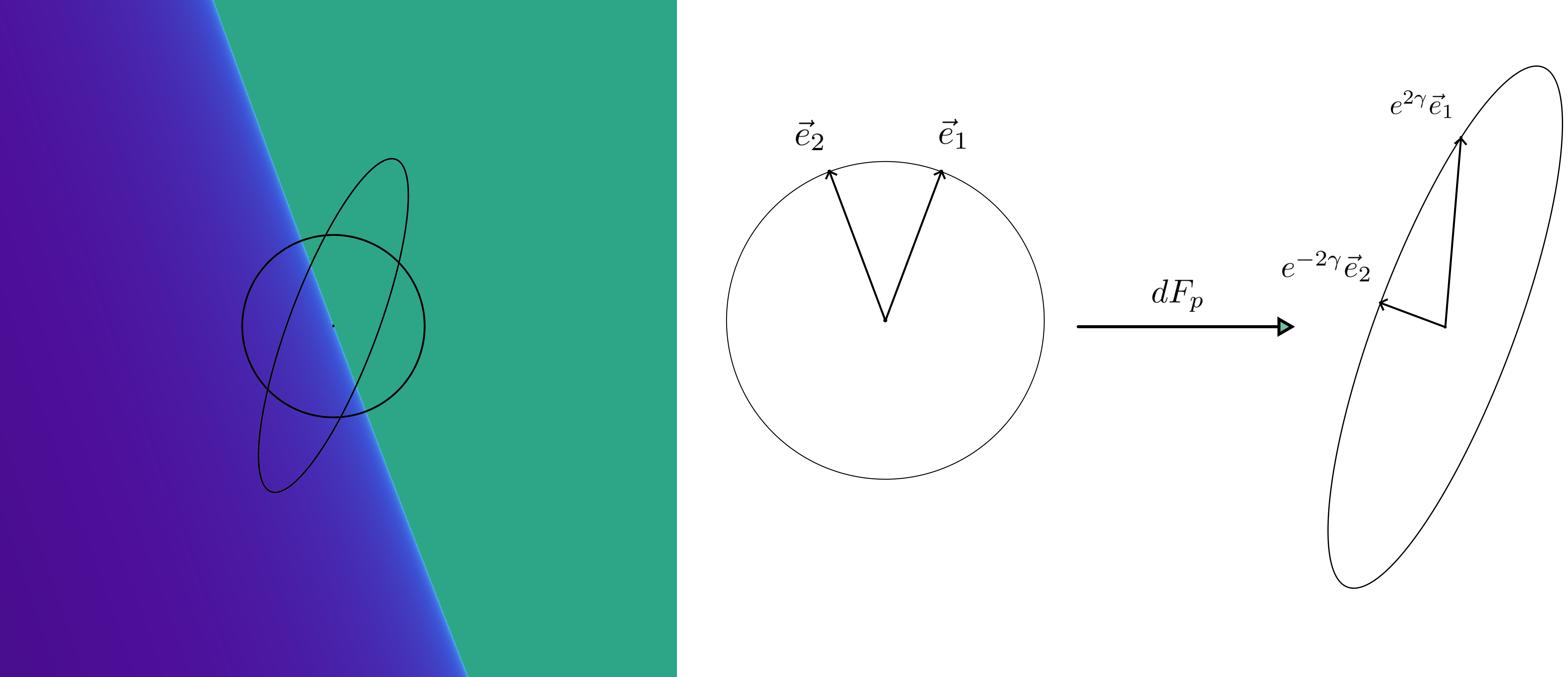}
    \caption{Linearized first-return dynamics near the hyperbolic fixed point associated with the unstable spherical photon orbit.
    \textbf{Left:} a small circle of initial conditions on the Poincaré section $\mathcal{S}$, centered at $(r, p_r) = (r_c, 0)$, and its image under the first-return map $F$ (shown as an ellipse), overlaid on the exit basin (green = escape, blue = capture).
    \textbf{Right:} schematic action of the differential $dF_p$ on infinitesimal deviations: a circular bundle is mapped to an ellipse.
    Deviations along the eigendirections $\vec{e}_1$ and $\vec{e}_2$ of $dF_p$ are stretched and contracted by factors $e^{\pm 2\gamma}$ along the unstable and stable directions, respectively.
    Symplecticity implies that the enclosed area is preserved under $dF_p$ before any visual rescaling introduced for legibility.}
    \label{fig:KerrFirstReturnDifferential}
\end{figure}

Next, we construct an exit basin in a neighborhood of the fixed point.
In Kerr, there are two possible exit channels: escape to large radius and capture by the black hole.
We terminate each trajectory when it reaches either $r_\text{capture} = 2.17 M$ or $r_\text{escape} = 100 M$, and record the exit channel.
Points on $\mathcal{S}$ are then colored green for escape and blue for capture, with the color gradient indicating the affine time required to reach the corresponding exit condition.
To illustrate how the linearized dynamics \eqref{eq:LinearizedDynamics} control the fine structure of these phase-space plots and the associated sensitivity to initial conditions, we evolve a family of deviations near the fixed point $(r = r_c, p_r = 0)$ up to first return.
To linear order, the evolution of an infinitesimal deviation $\delta \xi$ about a point $p$ is given by the (multiplicative) action of the differential of the return map, $dF_p$; see App.~\ref{app:PoincaréMaps} for details.
The eigenvalues and eigenvectors of $dF_p$ determine the invariant directions and the corresponding expansion and contraction factors after one return.
In the second frames of Figs.~\ref{fig:PhaseSpacePhotonRing} and \ref{fig:KerrReturnMap}, we zoom in on the fixed point of $F$---the unstable bound orbit---in the $(r, p_r)$ plane, taking one-sided widths $\pa{\frac{3}{100} M,\frac{3}{100} M}$.\footnote{The requisite zoom level was chosen by monitoring the maximum discrepancy between the linearized and nonlinear evolution of points on the circle at first return, and identifying a regime in which this error scales sublinearly.}
We then evolve a small circle of deviations to first return, obtaining an ellipse.
The deformation of this circle is very strong: the ratio of scale factors $\sigma_1/\sigma_2$ in the singular value decomposition of $dF_p$ at the fixed point is $\mathcal{O}(10^5)$.
For visual clarity, we therefore rescale the ellipse along its principal axes until it is legible.\footnote{Due to this rescaling, the ellipse is not drawn to scale and the apparent eigendirections are correspondingly distorted.}
The result is shown in Fig.~\ref{fig:KerrFirstReturnDifferential}.

Before the purely visual rescaling, the ellipse has the same area as the original circle, as required by symplecticity.
Moreover, one of the eigenvectors of $dF_p$ ($\vec{e}_2$, shown inset in the circle) lies along the separatrix, as it must: the separatrix is a contour of constant $\eta$ and consists of critically bound orbits, which remain asymptotically bound under each application of the return map.
The expansion and contraction along the eigendirections are given by $e^{\pm 2\gamma(r_c)}$, showing that the Lyapunov exponent evaluated at the critical radius $r_c$ controls the exponential sensitivity to initial conditions near the fixed point.
In this linearized regime, the self-similarity of the exit-basin coloring is likewise encoded in $dF_p$.
Indeed, a point on the initial circle and its image on the ellipse must share the same color, since both lie on the same phase-space trajectory and therefore have the same eventual fate: capture or escape to infinity.

To conclude, in the Kerr spacetime, one finds self-similarity in both the photon-ring substructure and the exit-basin coloring, together with exponential sensitivity to initial conditions near the hyperbolic fixed point, but not chaos, because the system remains integrable.

\section{Deformed Kerr: The Onset of Chaos}

In this section, we consider a particular class of stationary, axisymmetric deformations of the Kerr spacetime.
Using the geometric constructions and visualizations developed in the previous section, we identify which features of Kerr geodesic motion persist and which are lost under deformation.
In particular, the Carter constant is no longer conserved, the $r$ and $\theta$ dynamics no longer decouple, and integrability is broken.
The exponential sensitivity to initial conditions and self-similarity already present in Kerr are then supplemented by genuine chaotic behavior, which we illustrate through the fractal structure of the deformed exit basins.

\subsection{The Hartle--Thorne Spacetime}

We work with a particular subset of the Hartle--Thorne (HT) family of metrics \cite{Hartle:1967, Hartle:1968}.\footnote{\label{fn:HT}For readers familiar with this family, we set $\delta q = 1$ and $\delta m = 0$.}
The HT metric describes a stationary, axisymmetric spacetime that models the exterior geometry of a rotating, axially symmetric body.
It may be obtained as a perturbative solution of the Einstein equations by expanding in a small rotation parameter $\Omega$, with stationarity and axisymmetry imposed order by order.

The HT line element takes the form
\begin{subequations}
\begin{align}
    ds^2 &= - e^\nu \pa{1 + 2 h} \ed t^2 + e^\lambda \pa{1 + \frac{2 \mu}{r - 2m} } \ed r^2 \\
    &\phantom{=}\ + r^2 \pa{1 + 2k} \pa{\ed\theta^2 + \sin^2{\theta} \br{\ed \phi - \pa{\Omega - \omega} \ed t}^2} + \mathcal{O}\pa{\Omega^3},
\end{align}
\end{subequations}
where $\Omega$ is the angular velocity of the matter in the rotating body, and the metric potentials $h(r,\theta)$, $\mu(r,\theta)$, $k(r,\theta)$, and $\omega(r,\theta)$ are expanded in Legendre polynomials $P_\ell(\cos\theta)$ as
\begin{subequations}
\begin{gather}
    h(r, \theta) = h_0(r) + h_2(r) P_2, \qquad \mu(r, \theta) = \mu_0(r) + \mu_2(r) P_2, \\
    k(r, \theta) = k_2(r) P_2, \qquad \omega(r, \theta) = \omega_1(r) P_1'.
\end{gather}
\end{subequations}

We define $x \equiv r / M$ and $\chi \equiv a / M$.
For our chosen subset of HT metrics,\footref{fn:HT} we have
\begin{gather}
    e^\nu = e^{- \lambda} = 1 - \frac{2}{x}, \qquad
    \omega_1 = \Omega - \frac{2 \chi}{M x^3},
\end{gather}
and
\begin{subequations}
\label{eq:HT-Components}
\begin{align}
    \mu_0 &= - \frac{M \chi^2}{x^3}, \\
    \mu_2 &= - \frac{5 x M \chi^2}{16} \pa{1 - \frac{2}{x}}^2 \br{
        3x^2 \log \pa{1 - \frac{2}{x}} + \frac{2}{x} \frac{1 - 1 / x}{\pa{1 - 2 / x}^2} \pa{3x^2 - 6x - 2}} \notag \\
    &\phantom{=}\ - \frac{\chi^2}{x^2} \pa{1 - \frac{7}{x} + \frac{10}{x^2}}
    , \\
    h_0 &= \frac{\chi^2}{x^3 \pa{x - 2}}, \\
    h_2 &= \frac{5}{16} \chi^2 \pa{1 - \frac{2}{x}} \br{3x^2 \log \pa{1 - \frac{2}{x}} + \frac{2}{x} \frac{1 - 1 /x}{\pa{1 - 2 / x}^2} \pa{3x^2 - 6x - 2}} + \frac{\chi^2}{x^3} \pa{1 + \frac{1}{x}}, \\
    k_2 &= - \frac{\chi^2}{x^3} \pa{1 + \frac{2}{x}} - \frac{5}{8} \chi^2 \br{
        3 \pa{1 + x - \frac{2}{x} - 3 \pa{1 - \frac{x^2}{2}}} \log \pa{1 - \frac{2}{x}}
    }.  
\end{align}
\end{subequations}

As an aside, we note that this family of HT metrics exhibits stable light trapping \cite{Kostaros:2022}.
At first sight, this may seem to conflict with the proof of Kerr stability, since stable light trapping signals a geometric instability (or ``black-hole bomb''~\cite{Press:1972zz}): energy can accumulate in the trapping region and eventually backreact on the spacetime \cite{Cardoso:2014, Keir:2016, Cunha:2017}.
There is, however, no contradiction because the HT metric is not an exact solution of the Einstein field equations, whereas Kerr stability concerns dynamical perturbations of Kerr that solve the equations of general relativity.

\subsection{The Onset of Chaos}

To demonstrate the onset of chaos, we linearly interpolate between the Kerr and Hartle--Thorne spacetimes by introducing the metric
\begin{align}
    g_{\mu \nu}(\epsilon) = \epsilon g^{\rm HT}_{\mu \nu} + \left(1-\epsilon\right) g^{\rm K}_{\mu \nu},
\end{align}
where $g^{\rm K}_{\mu \nu}$ is the Kerr metric, $g^{\rm HT}_{\mu \nu}$ is the truncated Hartle--Thorne metric described above, and $\epsilon\in[0,1]$ parametrizes the interpolation.
We treat $g_{\mu \nu}(\epsilon)$ as a smooth family of stationary and axisymmetric metrics, not necessarily satisfying the field equations, that continuously deforms Kerr and breaks its integrability.
We will now repeat our prior analysis varying $\epsilon$.

The values of $E$, $L$, and $a$ are the same as in Figs.~\ref{fig:PhaseSpacePhotonRing} and \ref{fig:KerrReturnMap}.
We center the plotting window at $r = r_0(\epsilon)$ and $p_r = 0$, where
\begin{align}
    r_0(\epsilon) &= r_0^{\rm HT} + \frac{r_c - r_0^{\rm HT}}{1 + \mathrm{arctanh}\pa{\epsilon^2}},
\end{align}
with $r_0^{\rm HT} = 2.27 M$ and $r_c = 2.610235 M$ the Kerr critical radius corresponding to our chosen values of $E$ and $L$ (Sec.~\ref{subsec:PhaseSpaceRing}).
This recentering keeps the window near the boundary between escaping and captured photon trajectories as the spacetime is deformed.
The one-sided widths about this center are fixed at $(\Delta r, \Delta \hat{p}_r) = (3 \times 10^{-2} M, 3 \times 10^{-2} M)$, matching the dimensions of the middle panels in Figs.~\ref{fig:PhaseSpacePhotonRing} and \ref{fig:KerrReturnMap}.
Throughout, we also keep $r_\text{capture} = 2.17 M$ and $r_\text{escape} = 100 M$.

For nonzero $\epsilon$, some photons may reach neither $r_\text{capture}$ nor $r_\text{escape}$ within the allotted integration time.
We classify such trajectories as ``trapped.''
This trapping phenomenon was studied extensively in \cite{Kostaros:2022}.
It may signal an instability of the spacetime, as noted below Eqs.~\eqref{eq:HT-Components}.

As in Fig.~\ref{fig:KerrFirstReturnDifferential}, we color escaping photons green and captured photons blue; in addition, trapped photons are colored red.
For escaping and captured trajectories, the color gradient encodes the value of the affine parameter at which the corresponding exit condition is reached.
Finally, pixels corresponding to invalid initial conditions---that is, kinematically forbidden values of $r$ and $p_r$ for which the null constraint yields an imaginary $p_\theta$---are colored solid white.

\begin{figure}[htbp]
\centering

\begin{subfigure}{0.32\textwidth}
    \centering
    \includegraphics[width=\linewidth]{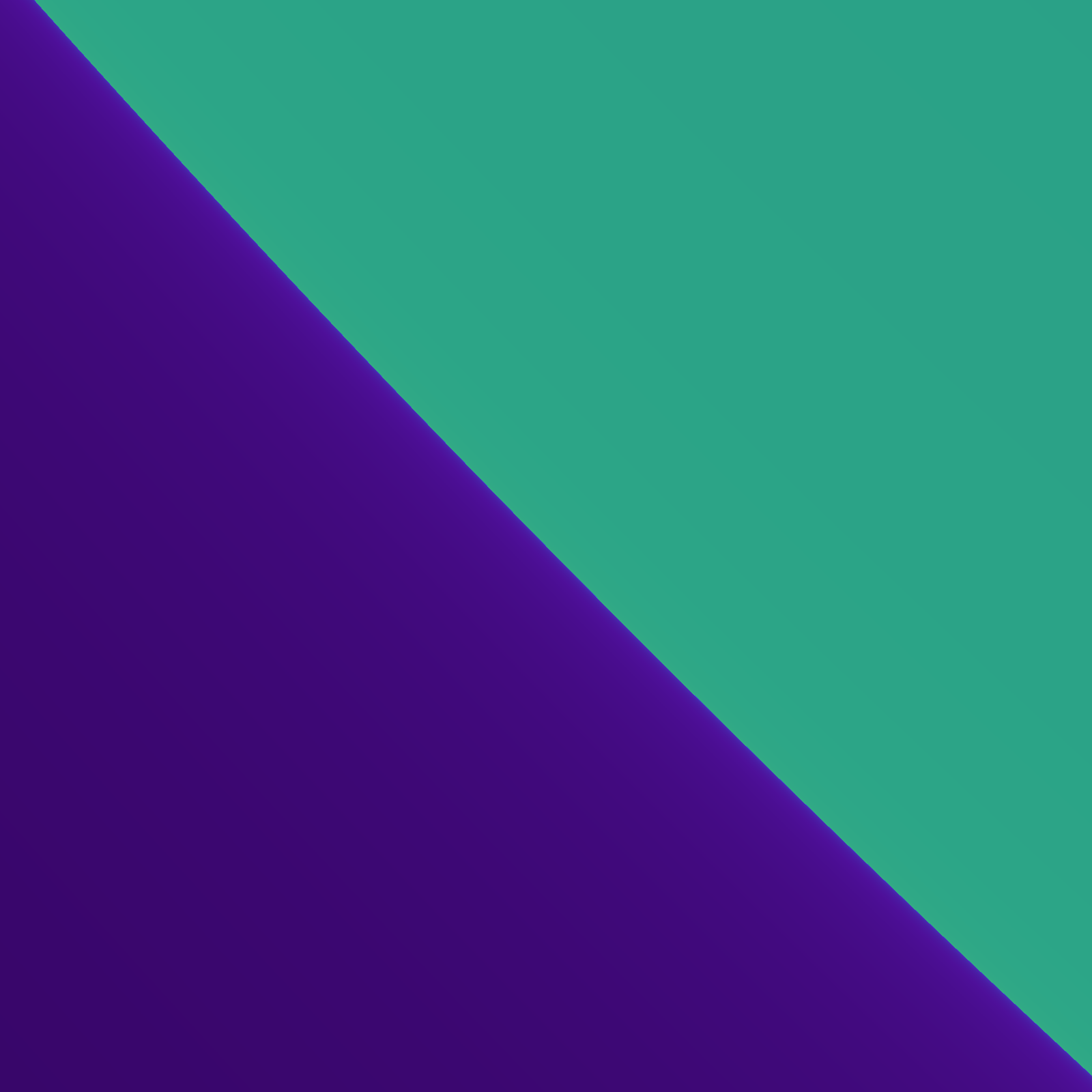}
    \caption{$N = 0$, $\epsilon = 0$}
\end{subfigure}
\hfill
\begin{subfigure}{0.32\textwidth}
    \centering
    \includegraphics[width=\linewidth]{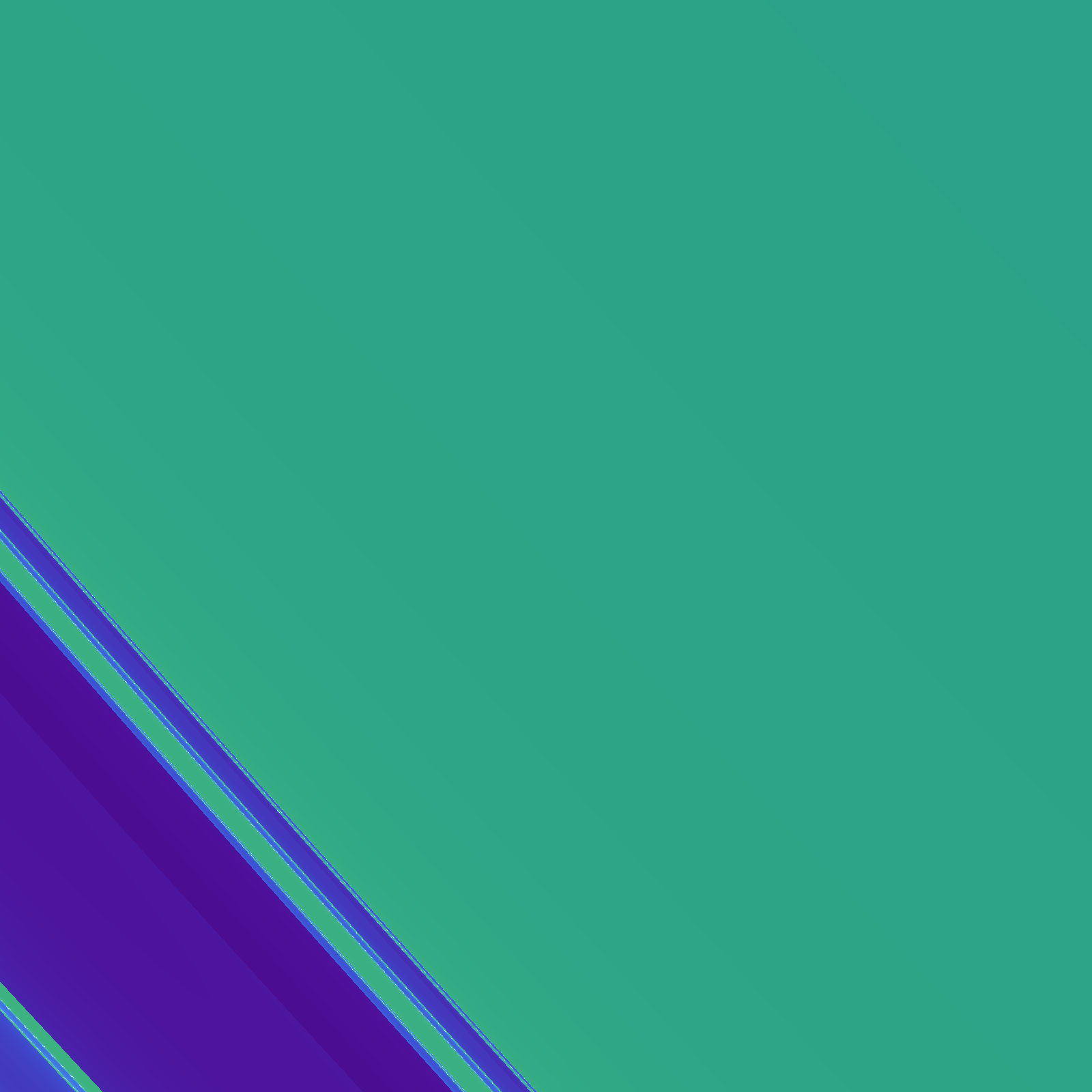}
    \caption{$N = 35$, $\epsilon = 0.9906$}
\end{subfigure}
\hfill
\begin{subfigure}{0.32\textwidth}
    \centering
    \includegraphics[width=\linewidth]{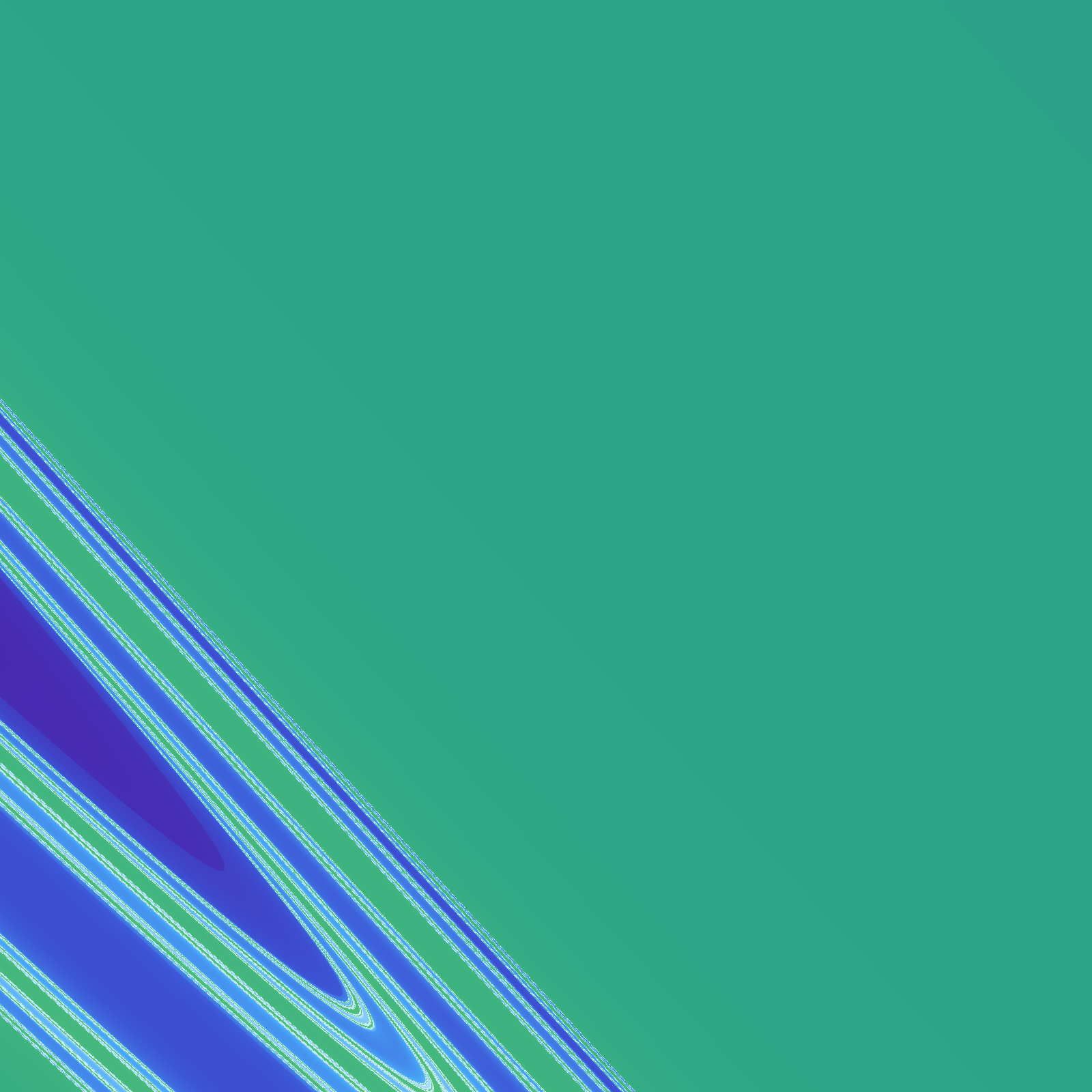}
    \caption{$N = 37$, $\epsilon = 0.9966$}
\end{subfigure}

\vspace{0.5em}
\begin{subfigure}{0.32\textwidth}
    \centering
    \includegraphics[width=\linewidth]{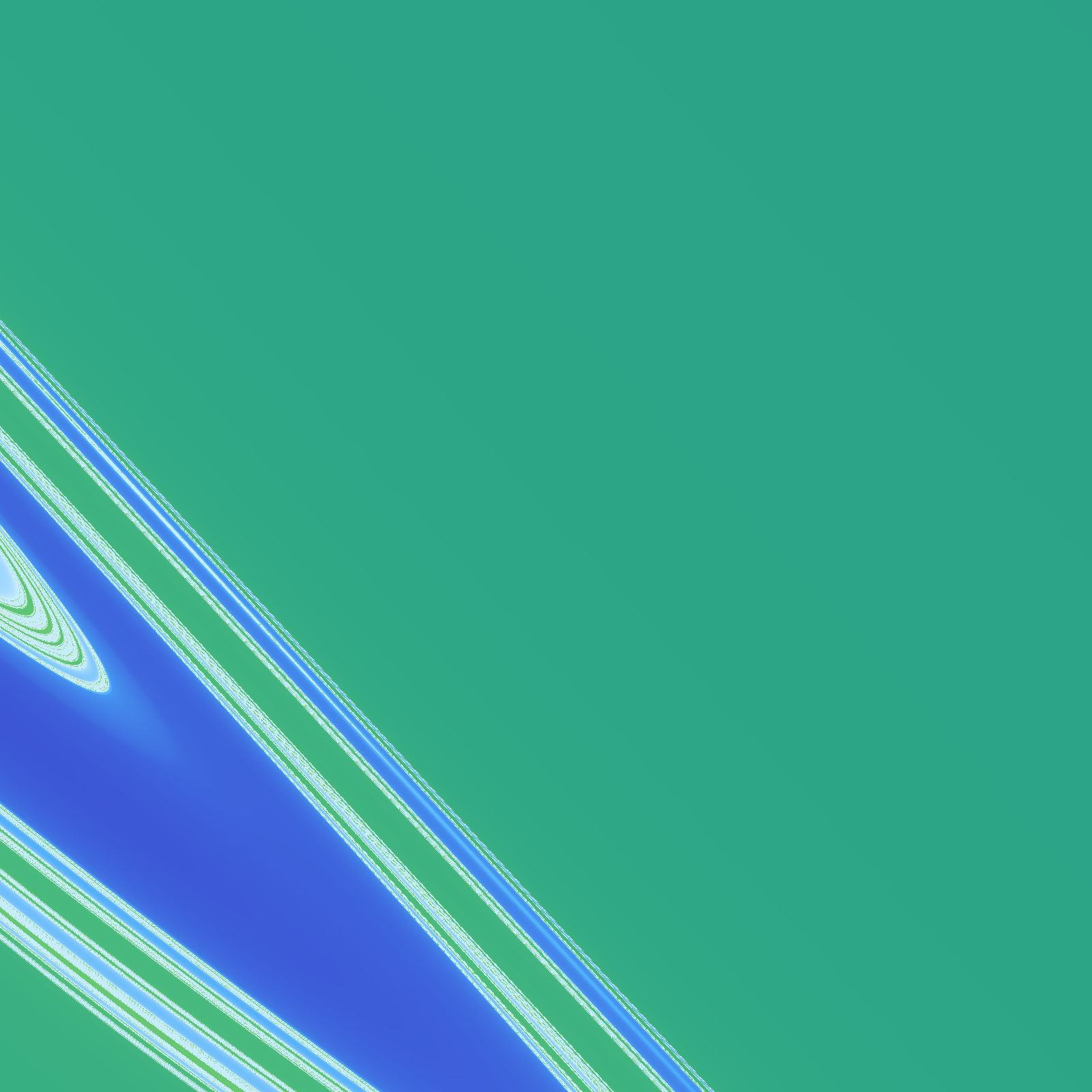}
    \caption{$N = 38$, $\epsilon = 0.9982$}
\end{subfigure}
\hfill
\begin{subfigure}{0.32\textwidth}
    \centering
    \includegraphics[width=\linewidth]{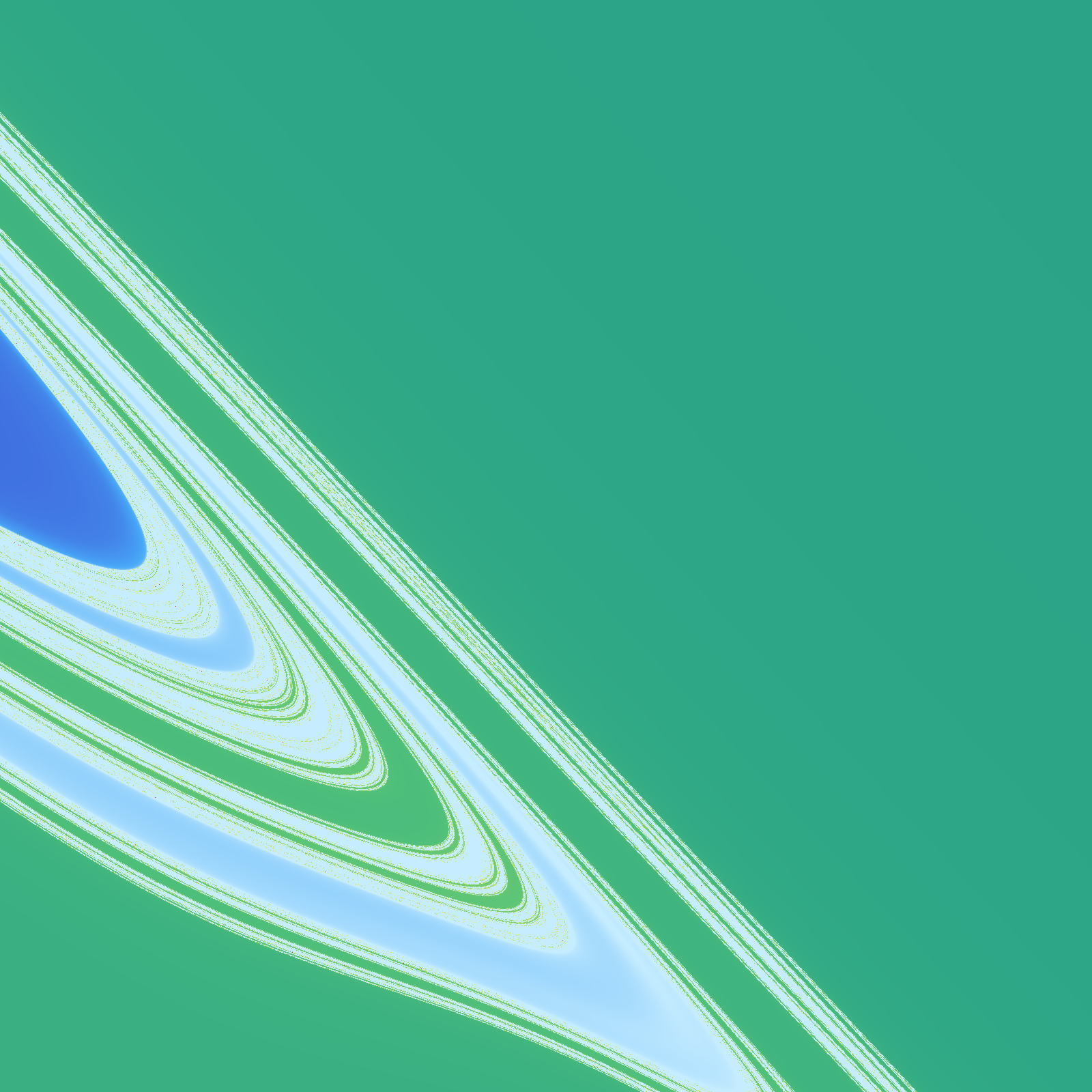}
    \caption{$N = 40$, $\epsilon = 0.999$}
\end{subfigure}
\hfill
\begin{subfigure}{0.32\textwidth}
    \centering
    \includegraphics[width=\linewidth]{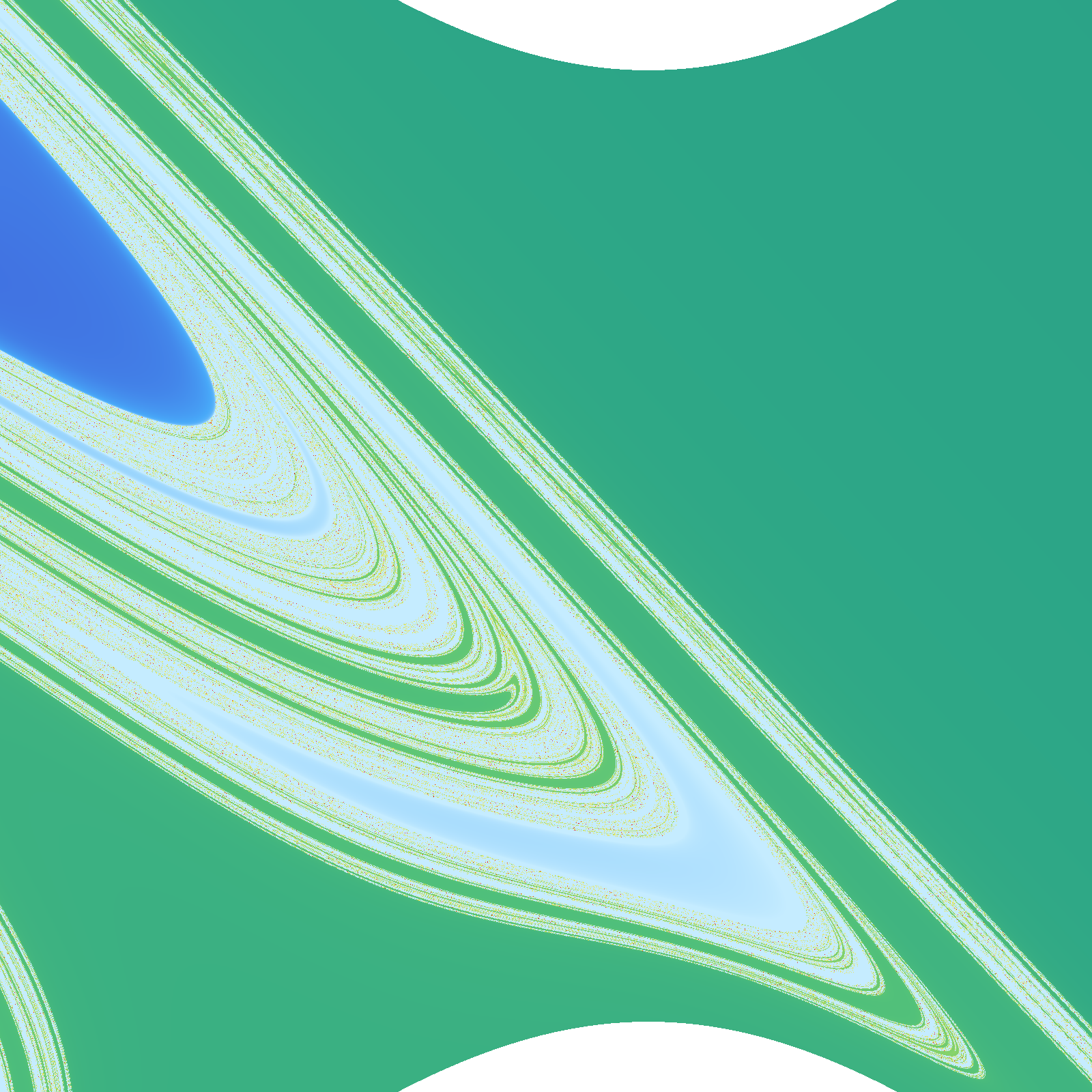}
    \caption{$N = 42$, $\epsilon = 0.9999$}
\end{subfigure}

\vspace{0.5em}
\hspace{0.25em}
\begin{subfigure}{0.40\textwidth}
    \centering
    \includegraphics[width=\linewidth]{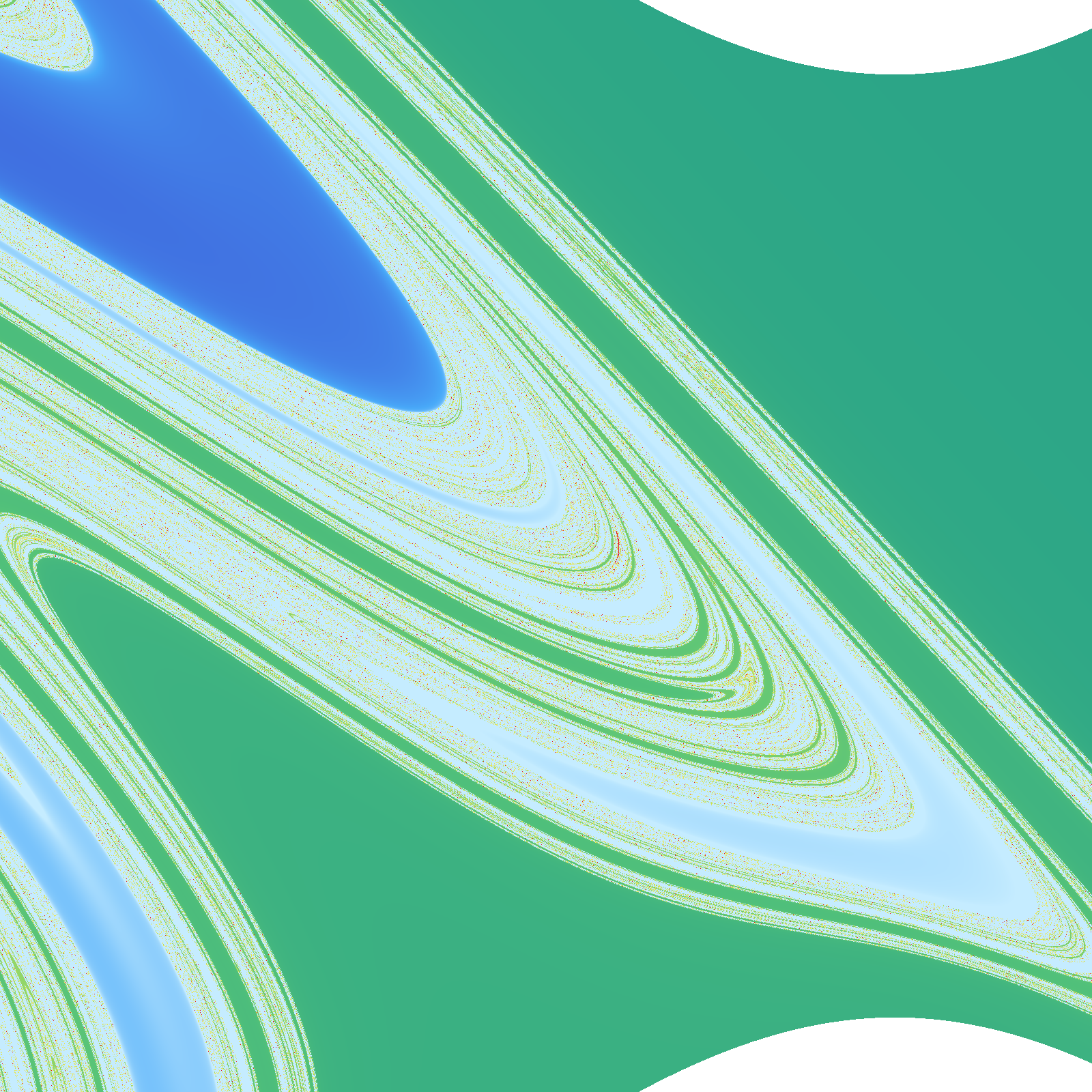}
    \caption{$N = 44$, $\epsilon = 0.999999$}
\end{subfigure}
\begin{subfigure}{0.40\textwidth}
    \centering
    \includegraphics[width=\linewidth]{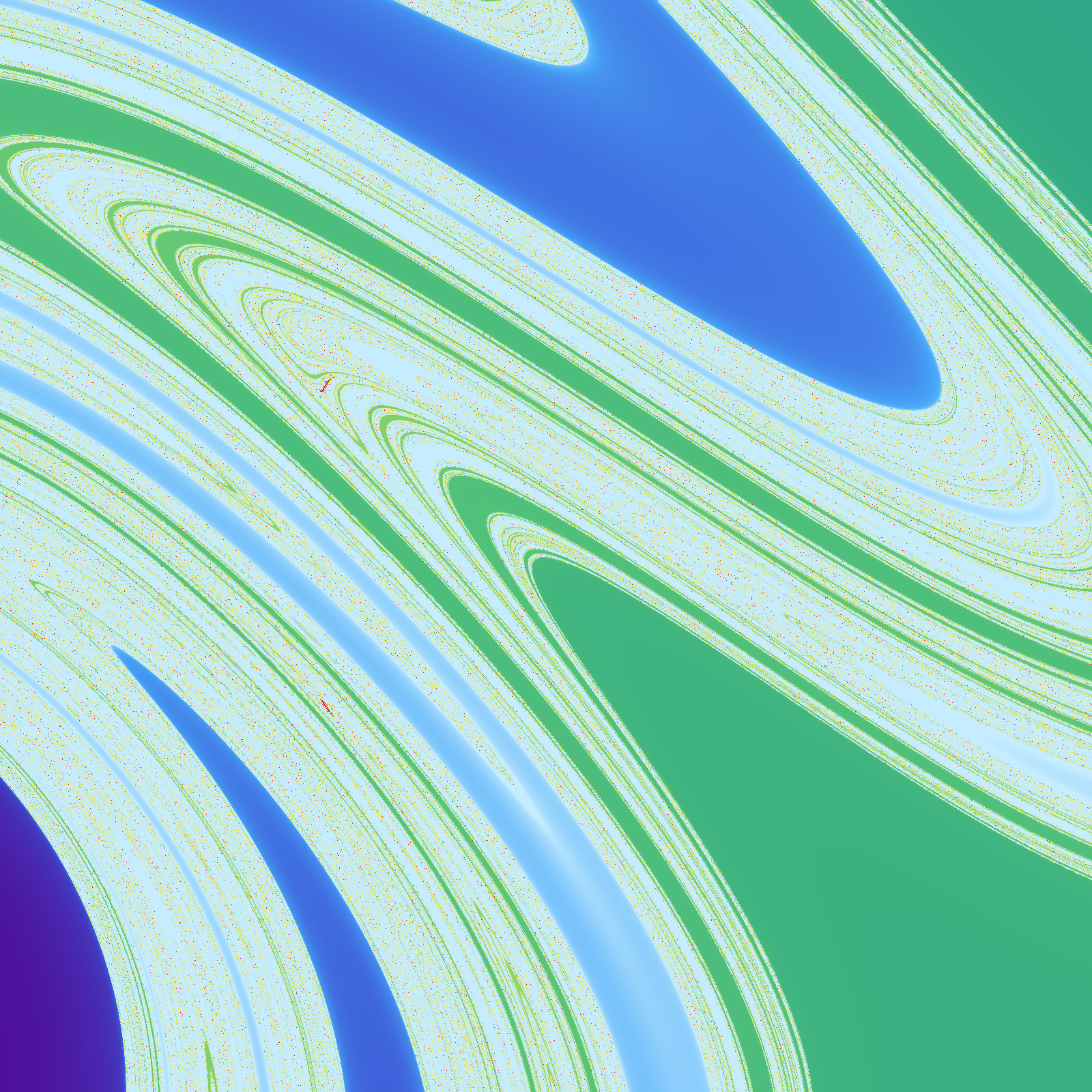}
    \caption{$N = 48$, $\epsilon = 0.99999999999$}
\end{subfigure}
\hspace{0.25em}

\caption{Representative frames from the interpolation between Kerr and Hartle--Thorne, illustrating the onset of chaos in the equatorial Poincaré section $\mathcal{S}$.
Each panel shows the exit basin in the $(r, p_r)$ plane for fixed $(E, L, a)$, with the plotting window recentered so as to track the boundary between capture and escape as $\epsilon$ varies.
Green pixels correspond to trajectories that reach $r_{\text{escape}}$, blue pixels to trajectories captured at $r_{\text{capture}}$, and the color intensity encodes the affine time required to reach the corresponding exit condition.
White denotes kinematically forbidden initial conditions, while red marks long-lived ``trapped'' trajectories that do not reach either exit channel within the allotted integration time.
For $\epsilon\ll1$ close to the Kerr limit, the basin boundary is a smooth separatrix; as $\epsilon\to1$ and the metric approaches the Hartle--Thorne end of the family, that boundary develops interleaved filaments and nested structures characteristic of a fractal basin boundary and sensitive dependence on initial conditions.}
\label{fig:MovieFrames}
\end{figure}

To generate the movie, we prescribe a map from the frame number $N$ (with $N=0,1,\dots,49$) to the deformation parameter $\epsilon$:
\begin{align}
    \epsilon(N) = \sqrt{\tanh\left(\frac{N}{50 - N}\right)}.
\end{align}
This choice concentrates most of the frames near $\epsilon=1$, where the fractal basin structure first becomes visible around $\epsilon \approx 0.99$ and then develops rapidly.
Representative frames from the movie are shown in Fig.~\ref{fig:MovieFrames}.
For the full animation, see \cite{OnsetOfChaosVideo}.

It is not until about frame $N=35$, corresponding to $\epsilon \approx 0.99064$, that the simple Kerr exit-basin structure---two connected regions separated by a diagonal interface---begins to give way to a complicated fractal pattern.
In integrable systems, phase space is foliated by invariant tori, and Kolmogorov--Arnold--Moser (KAM) theory describes how these tori break up as the system is perturbed into a chaotic regime.
A bound orbit in the photon shell is resonant when the ratio of angular frequencies $\Omega_\phi / \Omega_\theta$ is rational; such a resonance is especially strong when the reduced form $p / q$ of that rational number involves small integers.
Resonant orbits are closed and periodic, unlike nonresonant trajectories which densely fill their orbital shell.
In the standard KAM picture, resonant tori---especially low-order $p\!:\!q$ resonances---are the first places where chaos appears in the Poincaré section; see Sec.~3.2a of \cite{Lichtenberg:1992} for an illustration.
In the particular neighborhood of the separatrix associated with our Kerr critical orbit, which is not strongly resonant, the basin boundary remains smooth through most of the interpolation and becomes visibly fractal only once the deformation is sufficiently strong, at $\epsilon \gtrsim 0.99$.

\subsection{Self-similarity and First-Return Map}
\label{subsec:SelfSimilarity}

As Fig.~\ref{fig:MovieFrames} illustrates, once chaos sets in, the simple two-basin structure of Kerr is replaced by a complicated fractal pattern.
To probe the origin of this self-similarity, we again compute the differential of the return map, $dF_p$, at a point $p$ and study its action on infinitesimal deviations in a sufficiently small neighborhood of that point.

We focus on one deformed frame, $N=48$, and select a fixed point of the return map,
\begin{align}
    p = (r, p_r) = (2.295908 M, 0).
\end{align}
Around this point, we consider a small circle of deviations in a window with one-sided widths $(\Delta r, \Delta p_r) = (6 \times 10^{-5} M, 6 \times 10^{-5} M)$.
To turn this discrete action into a continuous animation, we choose a path through the space of symplectic matrices that begins at the identity and ends at $dF_p$, namely a curve $\gamma(s)$ such that each $\gamma(s)$ is symplectic and
\begin{align}
    \gamma(0)=\mathrm{Id}, \qquad \gamma(1)=dF_p.
\end{align}
Our specific choice of $\gamma(s)$ is defined in App.~\ref{app:SymplecticTrajectory}.
We stress that this interpolation is introduced purely for visualization: it is not physically preferred, but simply provides a continuous symplectic deformation from the identity to $dF_p$ that yields a smooth and informative animation.

\begin{figure}[t]
\centering

\begin{subfigure}{0.23\textwidth}
    \centering
    \includegraphics[width=\linewidth]{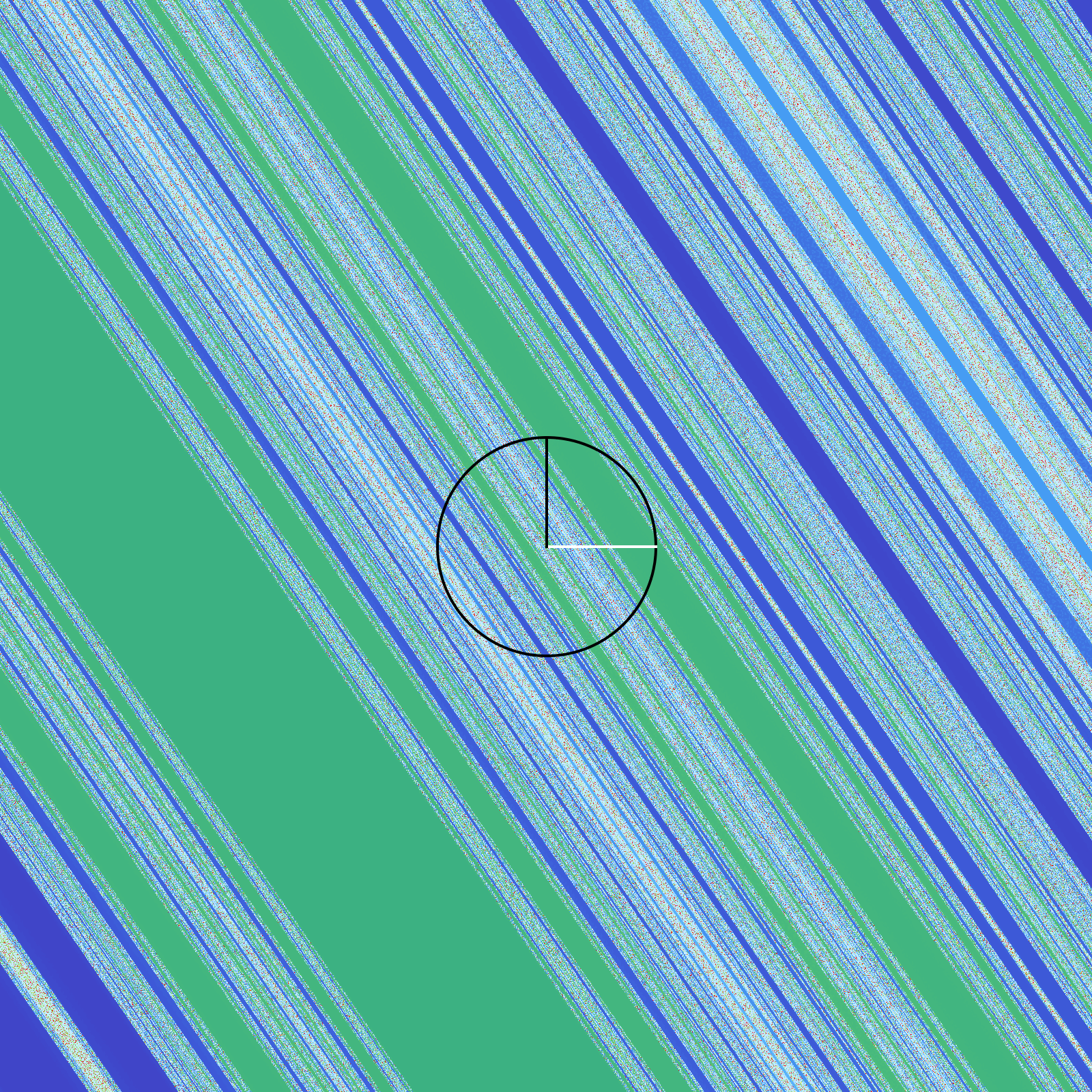}
\end{subfigure}
\hfill
\begin{subfigure}{0.23\textwidth}
    \centering
    \includegraphics[width=\linewidth]{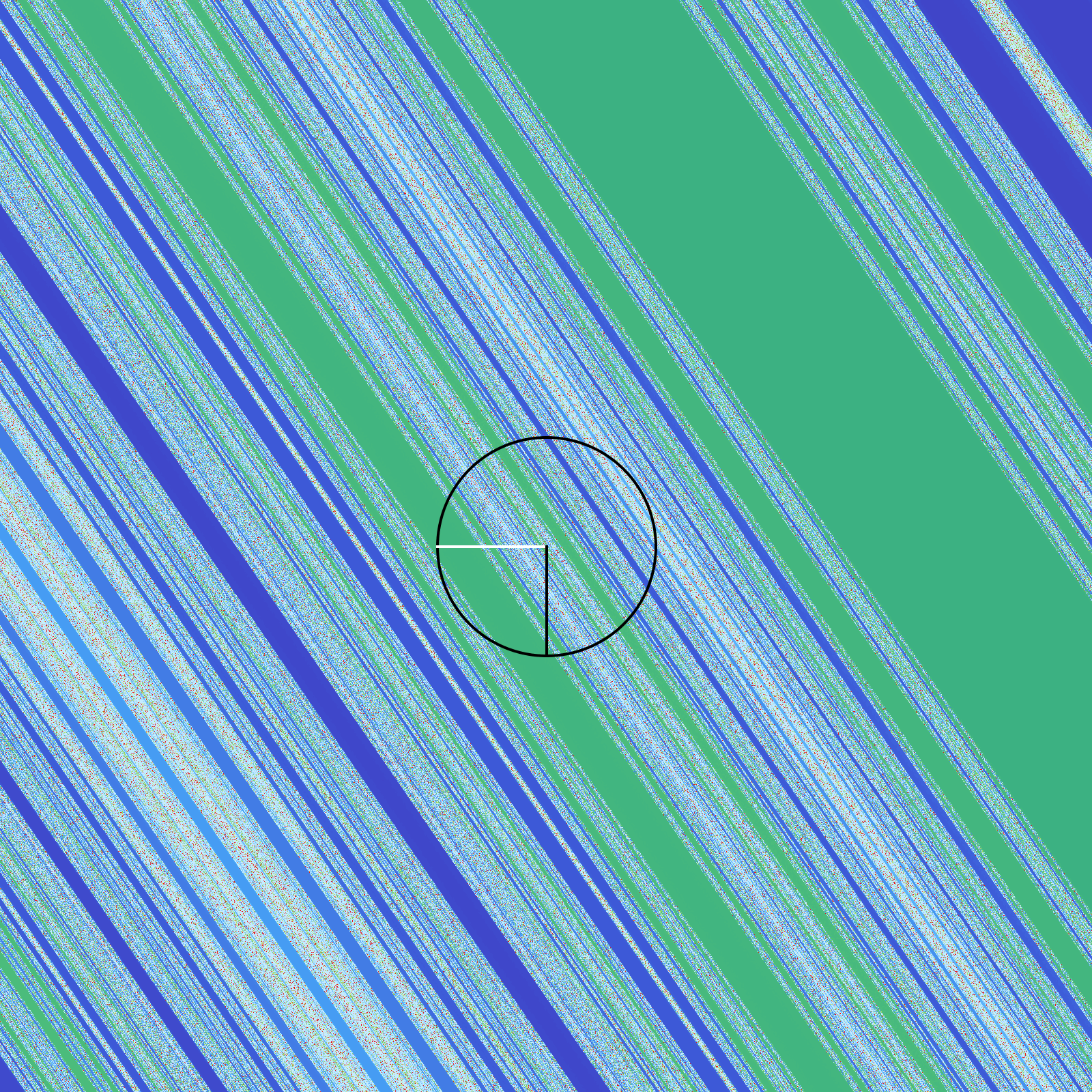}
\end{subfigure}
\hfill
\begin{subfigure}{0.23\textwidth}
    \centering
    \includegraphics[width=\linewidth]{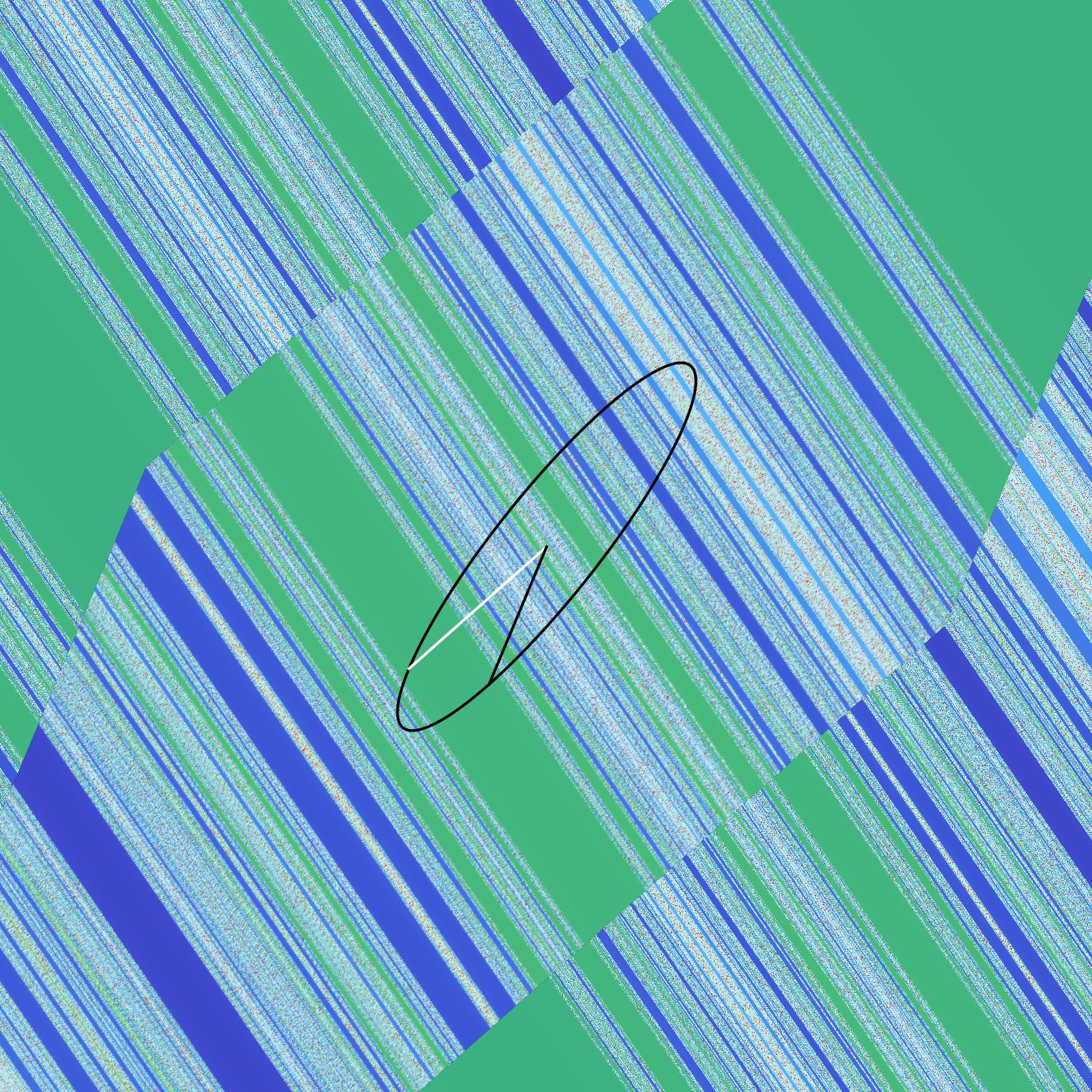}
\end{subfigure}
\hfill
\begin{subfigure}{0.23\textwidth}
    \centering
    \includegraphics[width=\linewidth]{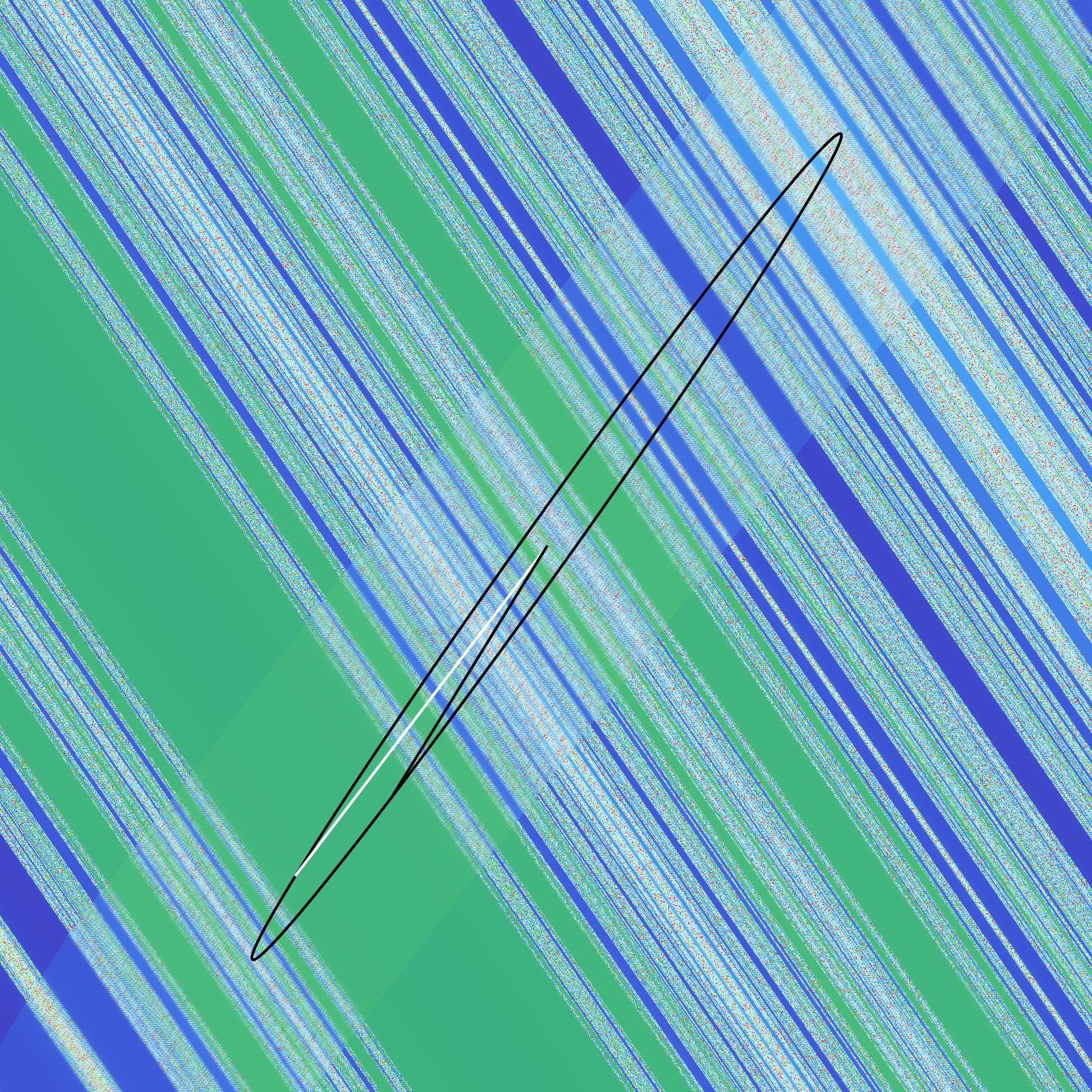}
\end{subfigure}

\caption{Linearized first-return action and emergent self-similarity in a chaotic exit basin.
The panels show representative frames of a continuous symplectic interpolation $\gamma(s) \in \mathsf{Sp}(2)$ from the identity to the differential of the first-return map $dF_p$, where $p$ is a fixed point of $F$.
The evolving overlay illustrates the progressive shearing and stretching of the basin texture under this local linear action.
Also shown is an infinitesimal circular bundle of initial conditions centered at $p$.
The agreement between the final overlay and the underlying basin coloring demonstrates that, sufficiently close to $p$, the fine filamentary structure of the basin boundary is encoded in the local linear dynamics governed by $dF_p$.}
\label{fig:ReturnMapSelfSimilarity}
\end{figure}

We then apply the matrices along this symplectic path continuously to the exit-basin image and superpose the transformed image on the original basin plot.
By the arguments of Sec.~\ref{subsec:PhaseSpaceRing}, the final transformed image should align with the underlying coloring pattern, since points related by the return map lie on the same trajectory and therefore share the same eventual fate.
In this way, the animation brings to light how the fine-grained fractal structure of the chaotic exit basin is encoded in the local linear dynamics.
Representative frames are shown in Fig.~\ref{fig:ReturnMapSelfSimilarity}. For the full animation, see \cite{SelfSimilarityVideo}.

\section{Discussion}

This work was motivated by a desire to clarify the precise relation between the black hole photon ring and the emergence of spacetime chaos.
In the geometry of a Kerr black hole, many familiar hallmarks of chaos are already present, including Lyapunov exponents and extreme sensitivity to initial conditions.
Yet geodesic motion in Kerr is completely integrable, and hence non-chaotic.
Here, we have found a satisfying resolution to this apparent tension: under generic deformations of the spacetime geometry---which are always present in the real world---the dynamics of photon-ring trajectories become genuinely chaotic.
Moreover, the photon shell is the locus in phase space where this transition first becomes manifest, and in particular, its resonant orbits witness the first onset of chaos.

We approached this problem visually because the phenomena themselves are so intricate: the folding, shearing, and stretching of escape basins; quasi-stable trapped matter; interleaved filaments; fractal chaos.
All of this is, in a sense, already implicit (albeit deeply buried) in the governing equations and in their numerical realization, but understanding the onset of chaos in this setting calls for a visual and geometric representation.

In a sense, our approach treats light as a frictionless fluid whose flow we follow into the chaotic regime.
This analogy may be sharpened by recalling a related idea from fluid dynamics, where the finite-time Lyapunov exponent is used to identify the ``skeleton'' of a flow, formed by so-called Lagrangian coherent structures (LCS) \cite{Haller:2015}.
Ridges (extrema) of the finite-time Lyapunov exponent field partition the flow into distinct domains across which air, water, or other transported substances separate.
Such structures arise in storm systems, at ocean-current boundaries, and in familiar patterns such as the von K\'arm\'an vortex street; they can often be seen directly in satellite images of cloud flow.
In this spirit, each bound photon orbit---and hence each corresponding orbital shell---may be viewed as an LCS ridge that separates photons on one side that plunge into the black hole from those on the other that escape to infinity.
The full collection of these ridges may then be regarded as composing the photon shell itself.

\begin{figure}[t]
    \centering
    \includegraphics[width=\textwidth]{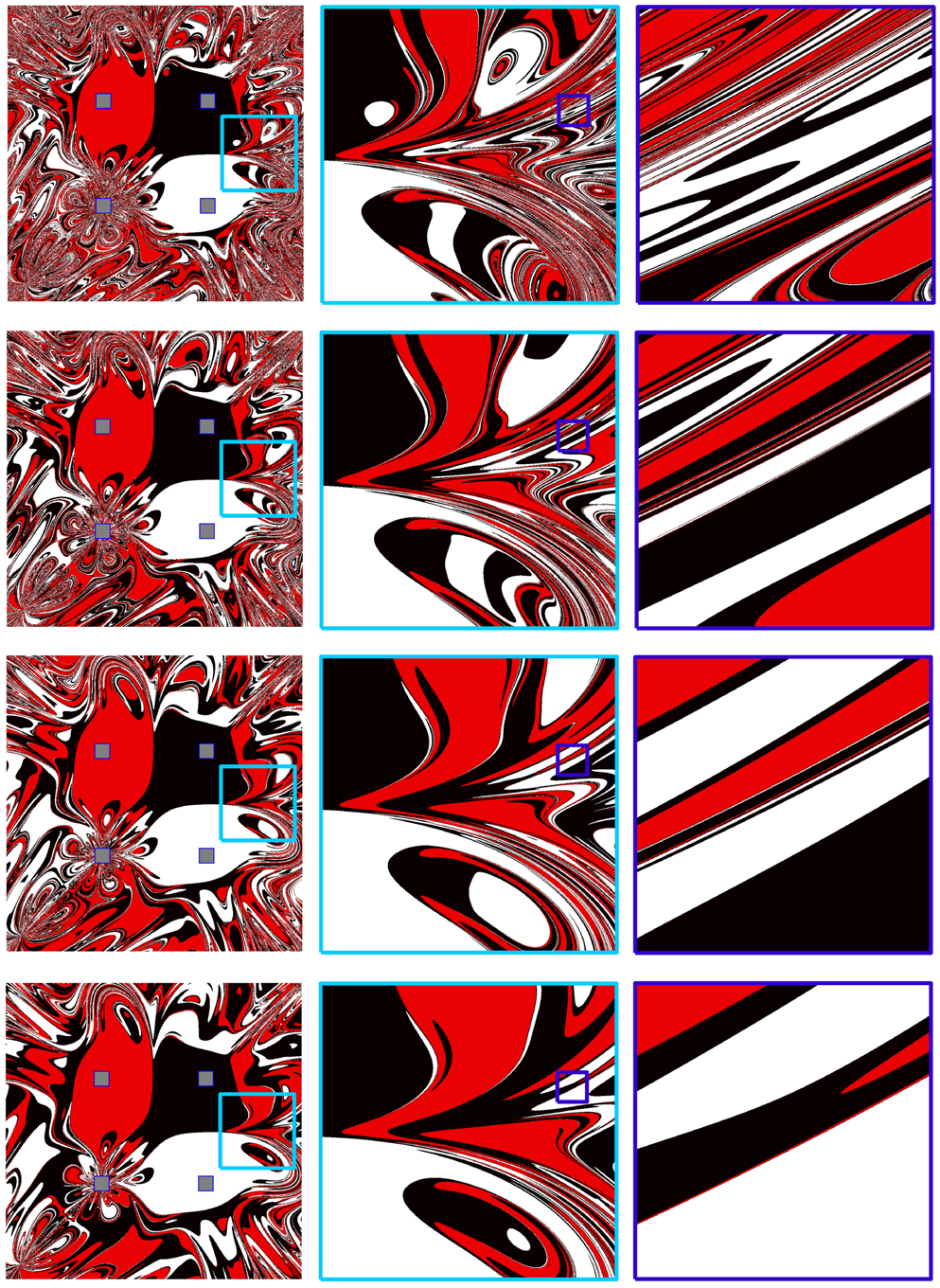}
    \caption{Reproduced (with permission) from Fig.~4 of \cite{Christian:2020}.
    Fractal basin structure of a suspended magnetic pendulum that swings above four fixed magnets placed on a table, with three attractive (labeled red, black, and white) and one repulsive.
    Initial positions of the pendulum above the table are colored according to the magnet over which the pendulum ultimately settles after tracing a complicated path.
    This manifestly self-similar structure is reminiscent of Figs.~\ref{fig:MovieFrames} and \ref{fig:ReturnMapSelfSimilarity}.}
    \label{fig:MagneticBasins}
\end{figure}

At a deeper level, the fractal basin structure is closely analogous to that of the magnetic pendulum: a suspended magnet that swings above, for instance, four fixed magnets placed on a table, with three attractive and one repulsive.
Once released from rest, the pendulum bob can wander unpredictably among the magnets, tracing a complicated path, before eventually settling over, or orbiting around, one of the three attractive ones.
Suppose these three magnets are labeled red, black, and white.
If one partitions the possible release points of the bob into small tiles and colors each tile according to the magnet over which the bob ultimately settles, the resulting plot reveals a striking feature: the differently colored regions (the basins of attraction) are separated by boundaries across which increasingly intricate, self-similar structure appears at every scale \cite{Grebogi:1983, McDonald:1985}.
Zooming in on such a boundary reveals smaller basins nested within larger ones, and further magnification uncovers still finer interleavings of the same kind.
This is beautifully illustrated in Fig.~4 of \cite{Christian:2020}, which we reproduce below in Fig.~\ref{fig:MagneticBasins}.
The resemblance to our Figs.~\ref{fig:MovieFrames} and \ref{fig:ReturnMapSelfSimilarity} is striking.

The deeper lesson is that whether one studies flows of light or the trajectories of pendula, one is ultimately studying physics unfolding in space and time.
Einstein taught us that light travels along geodesics; thus, when we examine the flow of light around a black hole, the resulting pattern traces the chaotic geometry of spacetime itself.

\vfill
\section*{Acknowledgments}

RB, TG, and AL were supported in part by NSF grant AST-2307888, the NSF CAREER award PHY-2340457, and the Simons Foundation grant SFI-MPS-BH-00012593-09.
PG was supported in part by the Gordon and Betty Moore Foundation grant 8273.01 and by the John Templeton Foundation grant 62286.
LCS was supported in part by NSF CAREER award PHY–2047382 and a Sloan Foundation Research Fellowship.
LCS thanks Samuel Lisi for strengthening his understanding of symplectic geometry.

\clearpage
\appendix

\section{Tangent Evolution}
\label{app:TangentEvolution}

We compute the evolution of tangent vectors in $\mathcal{S}$ to first return---equivalently, the differential $dF_p$ that maps the blue vector to the gray vector in Fig.~\ref{fig:PoincaréSection}---in three steps:
\begin{enumerate}
    \item Given $\delta r$ and $\delta p_r$, construct the embedded deviation vector $\delta \xi \in T_p \mathcal{M}$.
    \item Evolve $\delta \xi$ for a time $T(p)$ using the time-evolution operator to obtain $\delta \xi(T(p))$.
    \item Project $\delta \xi(T(p))$ back onto $\mathcal{S}$.
\end{enumerate}
We now describe these steps in turn.

Let $p \in \mathcal{S}$ be a point whose trajectory admits a first return.
The coordinates $(r, p_r)$ provide local coordinates on the section $\mathcal{S}$.
Let
\begin{align}
    \delta \xi = \delta r \pd_r + \delta \theta \pd_\theta 
    + \delta p_r \pd_{p_r} + \delta p_\theta \pd_{p_\theta}
\end{align}
be a tangent vector in $T_p\mathcal{M}$ written in the coordinate basis.
Given a choice of $\delta r$ and $\delta p_r$ specifying a deviation in $T_p \mathcal{S}$, the defining constraints of $\mathcal{S}$ determine the remaining components $\delta\theta$ and $\delta p_\theta$.
Indeed, recalling that $\mathcal{S} = \{ \theta = \pi / 2 \} \cap \{ \mathcal{H} = 0\}$, tangency to $\mathcal{S}$ requires
\begin{align}
    \delta \theta = 0, \qquad \delta \mathcal{H} = 0.
\end{align}
The first condition immediately fixes $\delta\theta$, while the second then determines $\delta p_\theta$.
In particular,
\begin{align}
    \delta \mathcal{H} = \frac{1}{2} \pa{p_r^2 \pd_r g^{rr} + p_\theta^2 \pd_r g^{\theta\theta} + 2 \pd_r V} \delta r + \pa{g^{rr} p_r} \delta p_r + \pa{g^{\theta\theta} p_\theta} \delta p_\theta,
\end{align}
where we used $\delta \theta = 0$.
Imposing $\delta \mathcal{H} = 0$ and solving for $\delta p_\theta$ then yields
\begin{align}
    \delta p_\theta = - \frac{g_{\theta \theta}}{p_\theta} \br{
        \frac{1}{2} \pa{p_r^2 \pd_r g^{rr} + p_\theta^2 \pd_r g^{\theta \theta} + 2 \pd_r V} \delta r + \pa{g^{rr} p_r} \delta p_r 
    }.
\end{align}
Thus the linear map that embeds $(\delta r, \delta p_r) \mapsto (\delta r, \delta\theta, \delta p_r, \delta p_\theta)$ into the full tangent space $T_p \mathcal{M}$ is
\begin{align}
    \text{Emb} =
    \begin{pmatrix}
        1 & 0 \\
        0 & 0 \\
        0 & 1 \\
        - \frac{g_{\theta \theta}}{2 p_\theta} \pa{p_r^2 \pd_r g^{rr} + p_\theta^2 \pd_r g^{\theta \theta} + 2 \pd_r V} 
        & - \frac{g_{\theta \theta} g^{rr} p_r}{p_\theta} 
    \end{pmatrix}.
\end{align}
(In fancier language, this is the differential of the inclusion map $\iota: \mathcal{S} \hookrightarrow \mathcal{M}$.)
The time-evolution operator is obtained by evolving the basis vectors $(\pd_r, \pd_\theta, \pd_{p_r}, \pd_{p_\theta})$ for a time $T(p)$ through the system \eqref{eq:LinearizedDynamics}, and then taking their images as the columns of the resulting operator.
We compute this operator numerically.
It is precisely the map $d \Phi_{T(p)}(p): T_p \mathcal{M} \to T_{q} \mathcal{M}$, where $q = \Phi_{T(p)}(p)$ is the first-return point of $p$ on $\mathcal{S}$.

Lastly, we must project back onto $\mathcal{S}$, since after evolution the deviation vector need not remain tangent to the section: it may acquire a component along the Hamiltonian flow $X_\mathcal{H}$.
As explained in App.~\ref{app:PoincaréMaps}, the evolved deviation can therefore be decomposed as
\begin{align}
    \delta \xi (T(p)) = \delta \xi_{\mathcal{S}} + \alpha X_\mathcal{H},
\end{align}
where $\alpha$ is a scalar and $\delta \xi_{\mathcal{S}}$ is the component tangent to $\mathcal{S}$.
Rearranging gives
\begin{align}
    \label{eq:ProjectedDeviation}
    \delta \xi_S = \delta \xi(T(p)) - \alpha X_\mathcal{H}.
\end{align}

Since the left-hand side lies in $\mathcal{S} = \Sigma \cap \mathcal{E}$, it must satisfy $\delta \theta = 0$, and hence so must the right-hand side.
This implies that
\begin{align}
    \delta \theta(T(p))  - \alpha X_\mathcal{H}^\theta = 0,
\end{align}
or equivalently, that
\begin{align}
    \alpha = \frac{\delta \theta}{g^{\theta \theta} p_\theta},
\end{align}
where all phase-space quantities are now understood to be evaluated at time $T(p)$.
Substituting this value of $\alpha$ into Eq.~\eqref{eq:ProjectedDeviation} yields the desired projected deviation.
In matrix form, the projection is implemented by
\begin{align}
    \text{Proj} = 
    \begin{pmatrix}
         1 & - \frac{X^r_\mathcal{H}}{g^{\theta \theta} p_\theta} & 0 & 0 \\
         0 & - \frac{X^{p_r}_\mathcal{H}}{g^{\theta \theta} p_\theta} & 1 & 0 
    \end{pmatrix}.
\end{align}

This completes the steps (1)--(3). The differential $dF_p: T_p \mathcal{S} \to T_q \mathcal{S}$ is 
\begin{align}
    dF_p = \text{Proj} \circ d\Phi_{T(p)}(p) \circ \text{Emb}.
\end{align}

\section{Poincaré Maps in Hamiltonian Systems}
\label{app:PoincaréMaps}

Here we review the coordinate-free formulation of Hamiltonian dynamics on a symplectic manifold, the geometric content of Liouville's theorem, and the reason discrete Poincaré maps inherit the area-preserving property of the underlying continuous flow.

Let $\mathcal{M}$ be a $2n$-dimensional manifold equipped with a closed, nondegenerate symplectic form $\omega$, so that $\ed\omega = 0$.
We follow the conventions of~\cite{MR2269239}, so that in local Darboux coordinates, the symplectic form takes the standard form $\omega = \sum_i \ed p_i \wedge \ed q^i$.
The Poisson bracket of any two functions is then $\{f,g\} = \omega^{-1}(\ed f, \ed g)$.
Any function $H$ determines an associated Hamiltonian vector field $X_H$ via
\begin{align}
    \omega({-}, X_{H}) &= \ed H,
\end{align}
or equivalently $X_{H} = \omega^{-1}({-},\ed H)$.
In particular, any scalar quantity evolves according to $\dot f = \{f,H\}$, so the phase-space coordinates $\xi^i$ satisfy
\begin{align}
    \label{eq:xi-dot}
    \frac{d}{dt} \xi^{i} = \{\xi^{i}, H\} = X_{H}^{i}.
\end{align}
We denote by $\Phi_{t}: \mathcal{M} \to \mathcal{M}$ the time-evolution map that advances a point $p\in\mathcal{M}$ by parameter time $t$ along the integral curve of $X_H$ through $p$.

More general objects, such as differential forms and vector fields, are Lie-dragged along $X_H$; for scalar functions this reduces to the evolution equation above.
The geometric statement of Liouville's theorem is that the symplectic form is preserved under this Hamiltonian flow:
\begin{align}
    \mathcal{L}_{X_H}\omega &= \ed \iota_{X_H}\omega + \iota_{X_H} \ed \omega \\
    &= \ed(-\ed H) + 0 = 0,
\end{align}
where $\iota_V$ denotes the interior product and we have used Cartan's magic formula in the first line.\clearpage

\noindent It follows that the area of an infinitesimal bundle of nearby trajectories (as defined by using $\omega$ to measure the signed area between any two tangent vectors) is conserved under the flow.
More generally, every exterior power of $\omega$ is preserved, $\mathcal{L}_{X_{H}} \omega^{k} = 0$, and in particular the top exterior power gives the conservation of phase-space volume for an infinitesimal bundle of nearby rays.

To make this more precise, let $\xi^i(t)$ be an integral curve of $X_H$, and let $\delta \xi^i(0)$ be an infinitesimal deviation vector.
The evolution of this deviation is obtained by linearizing Eq.~\eqref{eq:xi-dot} about the nonlinear trajectory $\xi^i(t)$.
This yields the linear system
\begin{align}
    \label{eq:DeviationVector}
    \frac{d}{dt} \delta\xi^{i} = \pa{\pd_{j} X_{H}^{i}} \delta \xi^{j} \equiv \mathcal{D}^{i}{}_{j} \delta\xi^{j},
\end{align}
where $\mathcal{D}^{i}{}_{j} = \pd_{j} X_{H}^{i}$ is a matrix-valued function of the phase-space point $\xi^k(t)$.
Thus, the matrix $I+\mathcal{D}(\xi(t))dt$ represents the infinitesimal linear map from the tangent space at $\xi(t)$ to the tangent space at $\xi(t+dt)$.

We can write a formal solution of this linear matrix ODE in terms of a path-ordered exponential:
\begin{align}
    \delta\xi(t) &= d\Phi_{t}\,\delta \xi(0), \\
    \delta\xi^{i}(t) &= \pa{ \mathrm{P} \exp \int_{0}^{t} \mathcal{D}(\xi(t')) dt'}^{i}_{\phantom{i}j}\, \delta \xi^{j}(0),
\end{align}
where $\mathrm{P}\exp\int$ is defined by the limit
\begin{align}
    d\Phi_{t} &= \mathrm{P}\exp\int_{0}^{t} \mathcal{D}(t') dt' \notag \\
    &\equiv \lim_{N \to \infty} \br{I + \mathcal{D}(t_{N}) \Delta t} \br{I + \mathcal{D}(t_{N-1}) \Delta t } \cdots \br{I + \mathcal{D}(t_{2}) \Delta t} \br{I + \mathcal{D}(t_{1}) \Delta t},
\end{align}
with $\Delta t = t/N$ and $t_k = k \Delta t$.
We also write $d\Phi_t : T_p \mathcal{M} \to T_{\Phi_t(p)} \mathcal{M}$, since this operator is precisely the differential (or pushforward) of the flow map $\Phi_t$.

The evolution of $\delta \xi$ may also be expressed in coordinate-independent geometric language.
Since $\frac{d}{dt} = X_H^i \pd_i$, Eq.~\eqref{eq:DeviationVector} is equivalent to $X_H^i\pd_i \delta\xi^j - \delta\xi^i\pd_i X_H^j = 0$, which states that the deviation vector is Lie-transported along the Hamiltonian flow:
\begin{align}
    \mathcal{L}_{X_H} \delta \xi = 0 .
\end{align}
Because $\mathcal{L}_{X_H} \omega = 0$ as well, the symplectic area of the parallelogram spanned by any two deviation vectors is preserved.
In particular, for two initial deviations $\delta \xi_1(0)$ and $\delta \xi_2(0)$,
\begin{subequations}
\begin{align}
    \mathcal{L}_{X_H} \pa{\omega(\delta \xi_{1}, \delta \xi_{2})} &= 0, \\
    \omega(\delta \xi_{1}(0), \delta \xi_{2}(0)) &= \omega(\delta \xi_{1}(t), \delta \xi_{2}(t))
    = \omega\pa{d\Phi_{t} \delta \xi_{1}(0), d\Phi_{t} \delta \xi_{2}(0)}.
\end{align}
\end{subequations}
A linear map $L$ satisfying $\omega(Lv, Lw)=\omega(v, w)$ is called symplectic.
Thus, $d\Phi_t$ is a symplectic transformation.

Before turning to Poincaré sections, let us briefly relate these matrices to Lyapunov exponents.
If the matrix $\mathcal{D}$ were constant, then the path-ordered exponential would reduce to the ordinary matrix exponential $\exp(t \mathcal{D})$.
In a basis that diagonalizes $\mathcal{D} = \mathrm{diag}(\lambda_{1}, \lambda_{2}, \ldots)$, the volume preservation of $d\Phi_{t}$ implies
\begin{align}
    \sum_i \lambda_i = 0.
\end{align}
In that basis, the components of $\delta \xi(t)$ along the eigenvectors grow or decay as $\exp(\lambda_i t)$, so the eigenvalues $\lambda_i$ may be identified with the local Lyapunov spectrum, and the largest one with the (local) Lyapunov exponent.
In general, however, $\mathcal{D}$ is not constant, so the relevant spectrum must instead be extracted from
\begin{align}
    \frac{1}{t} \log{d\Phi_t},
\end{align}
using the matrix logarithm.
Unfortunately, for generic points and finite evolution times, this spectrum is coordinate-dependent, because the matrix $(d\Phi_t)^i{}_j$ carries its lower index in the initial tangent space $T_p \mathcal{M}$ and its upper index in the final tangent space $T_q \mathcal{M}$.
At a fixed point, where $p = q$, this ambiguity disappears and the spectrum is coordinate-independent.

Now consider a constant-energy surface $\mathcal{E} = \{H = \mathrm{const.}\}$ in phase space, together with a second codimension-1 surface $\Sigma$ that intersects $\mathcal{E}$ transversally.
Their intersection $\mathcal{S} = \mathcal{E} \cap \Sigma$ is a $(2n-2)$-dimensional submanifold, called a Poincaré section.
The restriction of $\omega$ to $\mathcal{E}$ has rank only $2n-2$, since $\omega$ is alternating, and because $\omega({-},X_H) = \ed H$ annihilates vectors tangent to $\mathcal{E}$, the kernel of $\omega$ is precisely the span of the Hamiltonian vector field:
\begin{align}
    \ker\pa{\omega|_{\mathcal{E}}} = \mathrm{span}(X_H),
\end{align}
Let us denote by
\begin{align}
    \tilde{\omega} \equiv \omega|_{\mathcal{S}}
\end{align}
the restriction (or equivalently, pullback) of the symplectic form to $\mathcal{S}$.
We assume that $\tilde{\omega}$ has full rank $2n-2$ on $T\mathcal{S}$, so that it defines a symplectic structure on the submanifold $\mathcal{S}$.
We also assume that the Hamiltonian vector field $X_H$ is transverse to $\mathcal{S}$.
Under these assumptions, the tangent space to the energy surface admits a splitting
\begin{align}
    T\mathcal{E} = T\mathcal{S} \oplus \mathrm{span}(X_H)
    = T\mathcal{S} \oplus \ker\left(\omega|_{\mathcal{E}}\right).
\end{align}
Accordingly, every vector $v\in T_{\mathcal{S}} \mathcal{E}$ admits a unique decomposition
\begin{align}
    \label{eq:T_SE-Decomposition}
    v = v_{\parallel} X_H + v_{\mathcal{S}},
\end{align}
with $v_{\mathcal{S}}\in T\mathcal{S}$.
For any two vectors $v, w \in T_{\mathcal{S}} \mathcal{E}$, their signed area therefore satisfies
\begin{align}
    \label{eq:omega-tilde}
    \omega(v, w) = \tilde{\omega}(v_{\mathcal{S}}, w_{\mathcal{S}}).
\end{align}
In other words, measuring the symplectic area spanned by two vectors tangent to $\mathcal{E}$ using $\omega$ is entirely equivalent to measuring the symplectic area spanned by their components tangent to the Poincaré section using $\tilde{\omega}$.

Finally, we define the first-return map, also called the first-recurrence map or Poincaré map.
Let $\tilde{\mathcal{S}} \subseteq \mathcal{S}$ denote the subset of points whose trajectories return to $\mathcal{S}$.
The first-return map
\begin{align}
    F: \tilde{\mathcal{S}} \to \mathcal{S}
\end{align}
assigns to each point $p \in \tilde{\mathcal{S}}$ the first point at which the integral curve of $X_H$ through $p$ returns to $\mathcal{S}$.
Since the return time generally depends on the initial point, we write it as $T(p)$, so that
\begin{align}
    F(p) = \Phi_{T(p)}(p).
\end{align}

As noted above, the differential $d\Phi_t$ is a symplectic transformation, but it acts on the total tangent space:
\begin{align}
    d\Phi_t: T_p \mathcal{M} \to T_{\Phi_t(p)} \mathcal{M}.
\end{align}
Let us now examine its action on tangent vectors in $T_p \mathcal{S}$.
Fix some $p \in \mathcal{S}$ and an initial deviation vector $\delta \xi \in T_p \mathcal{S} \subset T_p \mathcal{E}$.
By construction, this vector has no component along $X_H$, so under the decomposition \eqref{eq:T_SE-Decomposition}, we simply have $\delta \xi = \delta \xi_{\mathcal{S}}$.
After first return, this vector is mapped to
\begin{subequations}
\begin{align}
    \delta \xi' &\equiv d\Phi_{T(p)} \delta \xi, \\
    \label{eq:delta-xi}
    \delta \xi' &= \delta \xi'_{\parallel} X_H + \delta \xi'_{\mathcal{S}},
\end{align}
\end{subequations}
and in general the image $\delta\xi'$ need not remain tangent to $\mathcal{S}$: it may acquire a component along $X_H$.
Geometrically, this reflects the fact that the nearby point $\xi+\delta\xi$ typically returns slightly earlier or later than $\xi$, as illustrated in Fig.~\ref{fig:PoincaréSection}.
However, by Eq.~\eqref{eq:omega-tilde}, only the component tangent to $\mathcal{S}$ contributes to the symplectic area within $T\mathcal{E}$.
Accordingly, given any two initial deviation vectors $\delta \xi_1$ and $\delta \xi_2$, we have
\begin{subequations}
\begin{align}
    \tilde{\omega}(\delta \xi_{1},\delta \xi_{2}) &= \omega(\delta \xi_{1}, \delta \xi_{2}), \\
    &= \omega\pa{d\Phi_{T(p)} \delta \xi_{1}, d\Phi_{T(p)} \delta \xi_{2}}
    = \omega\pa{\delta \xi_{1}', \delta \xi_{2}'}, \\
    &= \tilde{\omega}\pa{\delta \xi'_{1\mathcal{S}}, \delta \xi'_{2\mathcal{S}}}.
\end{align}
\end{subequations}
The first equality holds because the initial deviations lie in $T\mathcal{S}$; the second because $d\Phi_{T(p)}$ is symplectic; and the third because the component along $X_H$ lies in the kernel of $\omega|_{\mathcal{E}}$, so only the projections tangent to $\mathcal{S}$ contribute.
Thus, areas in $T\mathcal{S}$, measured by $\tilde{\omega}$, are preserved under the first return map.
Equivalently, the differential of the Poincaré map is symplectic with respect to $\tilde{\omega}$, and the first-return map is therefore area-preserving.

Equivalently, but more directly, one may work with the differential of the first-return map, $dF: T_{p}\mathcal{S} \to T_{F(p)}\mathcal{S}$.
By construction, the pushforward of a tangent vector $\delta\xi$ from $T_p\mathcal{S}$ will be another tangent vector in $T\mathcal{S}$, and it is in fact precisely $\delta \xi'_{\mathcal{S}}$ appearing in Eq.~\eqref{eq:delta-xi}.
One may compute this differential explicitly:
\begin{subequations}
\begin{align}
    \label{eq:dF-components}
    dF\ \delta\xi = d(\Phi_{T(p)})\ \delta\xi &= (d\Phi)_{T(p)} \delta \xi + (\delta \xi^{i} \pd_{i}T|_{p}) X_{H}, \\
    \label{eq:dF-abstract}
    dF &= (d\Phi)_{T(p)} + X_{H} \otimes dT.
\end{align}
\end{subequations}
The first term, $(d\Phi)_{T(p)} \delta \xi$, generally has a component transverse to $T\mathcal{S}$, but that component lies along $X_{H}$ and is exactly canceled by the second term, $(X_{H}\otimes dT) \delta \xi = X_{H}\,\delta t$, where $\delta t = \nabla_{\delta \xi} T$ is the time difference between the first return of $\xi + \delta \xi$ and $\xi$.

\section{Generating Coordinate-Time Evolution}
\label{app:TimeEvolution}

Given the Hamiltonian $\mathcal{H}$ in Eq.~\eqref{eq:Hamiltonian}, we seek a Hamiltonian $H = H(r, \theta, p_r, p_\theta) = H(x^a, p_a)$ whose Hamilton equations generate evolution with respect to Boyer--Lindquist coordinate time $t$:
\begin{align}
    \frac{dx^a}{dt} = \frac{\pd H}{\pd p_a},\qquad
    \frac{dp_a}{dt} = - \frac{\pd H}{\pd x^a}.
\end{align}
For an affine parameter $\tau$, we have
\begin{align}
    \frac{dx^a}{dt} = \frac{dx^a}{d \tau} \frac{d\tau}{dt} = \frac{\pd \mathcal{H} / \pd p_a }{\pd \mathcal{H} / \pd p_t},
\end{align}
and similarly,
\begin{align}
    \frac{dp_a}{dt} = \frac{dp_a}{d \tau} \frac{d\tau}{dt} = - \frac{\pd \mathcal{H} / \pd x^a }{\pd \mathcal{H} / \pd p_t}.
\end{align}
Therefore, we require
\begin{align}
    \label{eq:ReducedHamiltonianProperties}
    \frac{\pd H}{\pd x^a} = \frac{\pd \mathcal{H} / \pd x^a }{\pd \mathcal{H} / \pd p_t}, \qquad 
    \frac{\pd H}{\pd p_a} = \frac{\pd \mathcal{H} / \pd p_a }{\pd \mathcal{H} / \pd p_t}.
\end{align}
We claim that these relations are satisfied by taking $H = - p_t$, where $p_t$ is obtained by solving the null constraint $\mathcal{H} = 0$.
To see this, note that on the constraint surface $\mathcal{H} = 0$, we have
\begin{align}
    0 = \ed \mathcal{H} = \frac{\pd \mathcal{H}}{\pd x^a} \ed x^a + \frac{\pd \mathcal{H}}{\pd p_a} \ed p_a + \frac{\pd \mathcal{H}}{\pd p_t} \ed p_t.
\end{align}
Solving for $-\ed p_t$ gives
\begin{align}
    \ed H = -\ed p_t = \frac{1}{\pd \mathcal{H} / \pd p_t} \br{\frac{\pd \mathcal{H}}{\pd x^a} \ed x^a + \frac{\pd \mathcal{H}}{\pd p_a} \ed p_a}.
\end{align}
Since
\begin{align}
    \ed H = \frac{\pd H}{\pd x^a} \ed x^a + \frac{\pd H}{\pd p_a} \ed p_a,
\end{align}
comparing coefficients of $\ed x^a$ and $\ed p_a$ with the previous expression immediately yields Eq.~\eqref{eq:ReducedHamiltonianProperties}.

\section{Symplectic Trajectory}
\label{app:SymplecticTrajectory}

There are infinitely many paths through the space of $2\times2$ symplectic matrices that could be used to animate the discrete return map.
The simplest would be a one-parameter family of the form $\gamma(s) = \exp(s L)$ with $L\in \mathfrak{sp}(2)$ in the Lie algebra.
However, not every matrix $A \in \mathsf{Sp}(2)$ lies in the image of the exponential map.
In particular, the differential of the return map at the point $(r,p_r)=(2.295908M,0)$ considered in Sec.~\ref{subsec:SelfSimilarity} has matrix elements (in coordinates $(r, p_r)$)
\begin{align}
\label{eq:our-A}
    A =
    \begin{pmatrix}
        -2.2956842579391257 & -1.420290343108595 \\
        -3.0065432337475264 & -2.295682634831194
    \end{pmatrix},
\end{align}
and this matrix is not in the image of the exponential map from $\mathfrak{sp}(2)$.

Elements of $\mathsf{Sp}(2)$ are classified by their trace into three types: (i) elliptic, for which $|\Tr(A)|<2$; (ii) parabolic, for which $|\Tr(A)|=2$; and (iii) hyperbolic, for which $|\Tr(A)|>2$.
The image of the exponential map consists of all elements with $\Tr(A)>-2$, together with the special element $-I$; see, for example, Chapter~3, Exercise~22 of \cite{Hall:2015xtd}, which applies here via the exceptional isomorphism $\mathsf{SL}(2,\mathbb{R})\simeq\mathsf{Sp}(2)$.
As noted above, the matrix $A$ in Eq.~\eqref{eq:our-A} does not lie in this image.

We can circumvent this obstruction by factoring $A = (-A)(-I)$, where $-A$ \emph{does} lie in the image of the exponential map.
We may therefore view $A$ as the composition of a rotation by $\pi$ with the action of $-A$, and accordingly parameterize the interpolation by
\begin{align}
    \gamma(s) =
    \begin{cases}
    \begin{pmatrix}
        \cos \pa{2 s \pi} & - \sin \pa{2 s \pi} \\
        \sin \pa{2 s \pi} & \cos \pa{2 s \pi} 
    \end{pmatrix}
    & 0 \leq s \leq \frac{1}{2} \\
    \exp \pa{\pa{2 s - 1} \log \pa{ - A } }
    \begin{pmatrix}
        \cos \pa{\pi} & - \sin \pa{\pi} \\
        \sin \pa{\pi} & \cos \pa{\pi} 
    \end{pmatrix} & \frac{1}{2} \leq s \leq 1 \\
    \end{cases}.
\end{align}
Here, $\exp$ and $\log$ denote the matrix exponential and matrix logarithm, respectively.
The principal branch $L = \log(-A)$ lies in the real Lie algebra $\mathfrak{sp}(2)$, and hence each matrix $\exp(sL)$ is symplectic, providing a smooth interpolation from the identity to $-A$.
It follows immediately that $\gamma(0) = \mathrm{Id}$ and $\gamma(1) = A$, with every intermediate matrix along the path remaining symplectic.

\bibliographystyle{utphys}
\bibliography{Chaos}

\end{document}